\documentclass{aa}  
\pdfoutput=1
\usepackage{graphicx}
\usepackage[varg]{txfonts}
\usepackage[colorlinks]{hyperref}
\usepackage{natbib}
\bibpunct{(}{)}{;}{a}{}{,} %
\begin{document}
   \title{The EUV spectrum of the Sun: SOHO CDS NIS radiances during solar
     cycle 23}

   \author{
     V. Andretta\inst{1}
     \and
     G. Del Zanna\inst{2}
   }

   \institute{
     INAF, Osservatorio Astronomico di Capodimonte,
     Salita Moiariello, 16, 80141 Naples, Italy\\
     \email{andretta@oacn.inaf.it}
     \and
     DAMPT, Centre for Mathematical Sciences, 
     University of Cambridge,
     Wilberforce Road, Cambridge CB3 0WA, UK
   }

   \date{Received \ldots}

  \abstract{%
    For the first time, we present and discuss  EUV radiances  of the solar 
transition region (TR) and corona obtained during a solar cycle. 
The measurements were obtained with the
    SOHO/coronal diagnostic spectrometer (CDS) during the period from 1996 to
    2010.
We find that limb-brightening significantly affects any characterisation of 
the solar radiances. We present the limb-brightening function for the main lines 
and find that it does not change  measurably during the cycle.  
We confirm earlier findings that the radiance histogram of the cooler lines have a well
  defined, log-normal quiet-Sun component, although our results differ from previous ones.
The width of the lowest-radiance log-normal distribution is constant along the cycle.
Both the analysis of the centre-to-limb variation and of the radiance
statistical distribution point to a constant QS emission along solar cycle 23.
Lines formed above 1~MK are dramatically affected  by the presence of active
    regions, and    indeed, no ``quiet Sun'' region can be defined during periods of maximum
    activity.
 Much of the  irradiance variability in lines formed below 1.5~MK 
is due to a change in the emitting area.
 For hotter lines, the emitting area saturates to
  almost 100\% of full solar disk at the maximum of activity, while
  simultaneously the emission due to active regions increases by more than an
  order of magnitude.
We show that structures around active regions, sometimes referred to as
  dark halos or dark canopies,  are common and discuss their
  similarities and differences with coronal holes.  In particular, we show how
  they are well visible in TR lines, contrary to coronal holes.
}

  \titlerunning{SOHO CDS NIS radiances during solar cycle 23}
  \authorrunning{V.\ Andretta \& G.\ Del Zanna}
  \maketitle

\defcitealias{DelZanna-etal:10}{Paper~I}
\defcitealias{DelZanna-Andretta:11}{Paper~II}

\section{Introduction}\label{sec:intro}

The  solar \& heliospheric observatory (SOHO) 
 coronal diagnostics spectrometer  
\citep[CDS;][]{Harrison-etal:95}
has been performing routine radiance measurements of the extreme ultraviolet (EUV) Sun since 
1996; hence, it offers a unique opportunity to study the 
characteristics of the solar radiance along a solar cycle in 
many spectral lines, which are formed in different layers of the solar 
atmosphere.
With a few minor gaps, CDS covers the important 150--800~\AA\ EUV spectral
range, of which the 308--379 and 513--633~\AA\ are covered at first order by
the two normal incidence spectrometer spectral channels (NIS 1 and 2, respectively), observed at
  a spectral resolution of about $\sim 0.5$~\AA\ or better.

The main aim of this paper is to characterise the solar EUV
radiances during an entire  cycle (cycle 23) for the first time.  
Previously, the only detailed study on EUV radiances
was based on data taken 
with full-Sun scans of the SOHO  
solar ultraviolet measurements of emitted radiation
(SUMER, \citealt{Wilhelm-etal:95}).
Such a study
provided, however, radiance and irradiance measurements in a 
few spectral lines only and for a limited time interval during the solar
cycle during minimum solar conditions in 1996 \citep{Wilhelm-etal:98}. 

This paper is a part of a programme  of studying the 
EUV spectral  radiance, $I$,
and irradiance, $F$, over the solar cycle.
Preliminary results were presented in 
\citet{DelZanna-etal:05} and \citet{DelZanna-Andretta:06}.
A detailed in-flight CDS calibration was
presented by \cite{DelZanna-etal:01_cdscal}, where responsivities for all 
the nine  channels were provided.
A detailed study of the long-term drop in sensitivity of the 
NIS channels during the 1998-2010 period 
was presented in \cite{DelZanna-etal:10} (henceforth:
\citetalias{DelZanna-etal:10}), where the \cite{DelZanna-etal:01_cdscal}
responsivities were slightly revised. 
The new calibration allowed the first measurements of the 
EUV spectral irradiance along a solar cycle, as presented in 
\cite{DelZanna-Andretta:11} (henceforth: \citetalias{DelZanna-Andretta:11}).
This latter paper discussed all hystorical records in detail
and obtained a revised calibration for the important 
\ion{He}{ii} 304~\AA\ line, as observed in second order with the 
NIS 2 channel, and the most prominent line in the EUV.
Large discrepancies in the \ion{He}{ii} 304~\AA\ radiances
have been present in the literature, however recent 
comparisons of our measurements with those obtained 
with the EUNIS (EUV normal incidence spectrometer) rocket flights 
(see, e.g., \citealt{Jordan-Brosius:07}) has uncovered 
errors in the EUNIS and CDS analysis software, and now
excellent agreement (within a relative 10\%)
 is found between our measurements
and those obtained with EUNIS (see \citealt{WangT-etal:11})
for the \ion{He}{ii} 304~\AA\ line and other
prominent EUV lines. 
Good agreement (see \citetalias{DelZanna-Andretta:11}) was also found between our CDS
measurements and those obtained with a sounding rocket 
flight in 2008  for a prototype of the
the solar dynamics observatory (SDO) 
 extreme ultraviolet variability experiment (EVE, see 
\citealt{Woods-etal:12}).
These favourable comparisons validate our long-term
calibration and therefore allow
 a quantitative description 
of the distribution of 
solar radiances along a solar cycle to be obtained from the CDS measurements for the first time.
This has important implications in itself but also allows us to 
understand where the variability in the solar irradiance
occurs, hence, to interpret X-UV irradiance measurements of the Sun as a star.

A new remarkable finding of \citetalias{DelZanna-Andretta:11} was the little 
variability of the irradiances of lines formed at all
transition-region (TR) temperatures
(e.g. those from \ion{O}{iii}, \ion{O}{iv}, \ion{O}{v}), 
as observed with SOHO/CDS.
This was also shown to be evident in the data from 
the NASA  thermosphere ionosphere mesosphere energetics dynamics (TIMED)
 solar EUV experiment (SEE) EUV grating spectrograph (EGS) 
\citep{Woods-etal:05}, which has data available since 2002.
This issue is particularly important, considering that,
as shown in \citetalias{DelZanna-Andretta:11}, the mostly used models 
 used in Earth's upper atmosphere research 
(e.g.  HEUVAC, \citealt{Richards-etal:06},
SIP,  \citealt{Tobiska-etal:08}, 
and the older H81 of \citealt{Hinteregger-etal:81})
overestimate the irradiances of TR lines by large factors,
which, in turn, represent a dominant component 
of the solar UV irradiances.

The scheme of this work is the following: We summarise the observations used
and the data reduction procedures in Sec.~\ref{sec:obs}; we discuss some
general properties of the statistical distributions of EUV radiances in
Sec.~\ref{sec:disc:hist} (including a general description of the various
contributing solar features); a more detailed analysis of the properties of
the quiet Sun radiances is given in Sec.~\ref{sec:disc:qs} (including
measurements of centre-to-limb variations and of histogram mean properties).
Finally, we address the above issues concerning the variability of the EUV
solar irradiance in Sec.~\ref{sec:disc:irr}.

\section{Observations and data analysis}\label{sec:obs}

Since the beginning of the SOHO mission, the CDS spectrograph has been used to
monitor daily the central meridian of the Sun in a few selected spectral
lines.  The scan of the central meridian is accomplished through a mosaic of
nine spatial rasters that each cover $240\times 240\arcsec$ on the Sun 
with a sequence of observations comprising the CDS study called SYNOP 
using the 2\arcsec\ slit (the majority of times).  There are some
variants of that study but all  include the lines  \ion{He}{i} 584~\AA,
\ion{O}{v} 630~\AA, \ion{Mg}{ix}~368~\AA,  and \ion{Mg}{x} 625~\AA.
We  selected a (relatively small) sample of synoptic observations spanning
the 13 years of SOHO operations, which tries to uniformly cover the whole period
but  also  avoids days when strong active regions
dominated  at the same time.

In addition, the study called USUN has been used since 1998 to scan the full
solar disk with a mosaic of 69 rasters that each cover
$240\times 240\arcsec$ (see, e.g., \citealt{Thompson-Brekke:00} 
and \citetalias{DelZanna-etal:10} for details).
The movement of the slit is ``sparse'' in the sense that 
the 4\arcsec\ slit is moved in larger steps  than its width. 
The radiance of the Sun is therefore subsampled by about a 
factor of 6 until to the end of 2002 (step size of 24.38\arcsec) and
  by a factor of 4 afterwards (step size of 16.25\arcsec). 
Exposures are also binned on board along the slit (for an equivalent pixel
size of 13.44\arcsec) to 
increase the count rates and reduce the telemetry load.
Contrary to the SYNOP studies, the USUN
studies record the full spectrum in the NIS wavelength range.

We applied standard processing for the  de-biasing, flat-fielding,
and  2\arcsec\ slit burn-in
using the CDS routine VDS\_CALIB, but
we developed custom-written analysis software for all the processing 
(cosmic rays removal, mosaicing the rasters, long-term calibration, etc.).
All the lines in the  spectra  were fitted with 
simple Gaussian profiles
in the case of data prior the loss of contact of SOHO 
and with properly broadened profiles 
(with software developed by W.T.Thompson%
\footnote{CDS Software Note \#53:\\
  \texttt{http://solar.bnsc.rl.ac.uk/swnotes/cds\_swnote\_53.pdf}}%
)
afterwards.
The largest uncertainty in the fit is the location of the background (the
scattered light), in particular for the NIS 1 channel. The fitting in the NIS
channels was done on the spectra in photon events, where the scattered light
``background'' component is more constant.  This uncertainty is more evident in
SYNOP spectra due to detector windowing, which limits the possibility of
estimating the background outside the line profile.

All the calibration factors have been applied after line fitting.
They mostly include  the wavelength-dependent 
 long-term corrections for the  drop in  responsivity as detailed
in \citetalias{DelZanna-etal:10} (extended in \citetalias{DelZanna-Andretta:11}), and the \cite{DelZanna-etal:01_cdscal} responsivities,
which are modified as described in \citetalias{DelZanna-etal:10} (and for the \ion{He}{ii}~304~\AA\
as described in \citetalias{DelZanna-Andretta:11}).

\section{Discussion}\label{sec:disc}

The SYNOP and USUN studies are in some respects complementary: SYNOP studies
are characterised by a longer time coverage (from 1996 to present) and higher
spatial resolution (pixel size of 2$\times$1.68\arcsec, while resolution
elements in USUN scans are of 24.38$\times$13.44\arcsec\ or
16.25$\times$13.44\arcsec) but are limited to a strip around the central
meridian and to a few spectral lines.  The USUN studies, on the other hand, cover
the full solar disk and the full NIS spectral range, with a typically higher S/N
(signal-to-noise) ratio than SYNOP observations.
In our analysis, we mainly exploit the USUN studies, but we also use the SYNOP
mosaics to validate and extend the USUN data.

\subsection{Statistical distribution of solar EUV radiances}\label{sec:disc:hist}

\subsubsection{Comparison of SYNOP and USUN statistics}\label{sec:disc:hist:comp}

\begin{figure*}[!htbp]
  \centering
  \includegraphics[width=0.475\linewidth]%
  {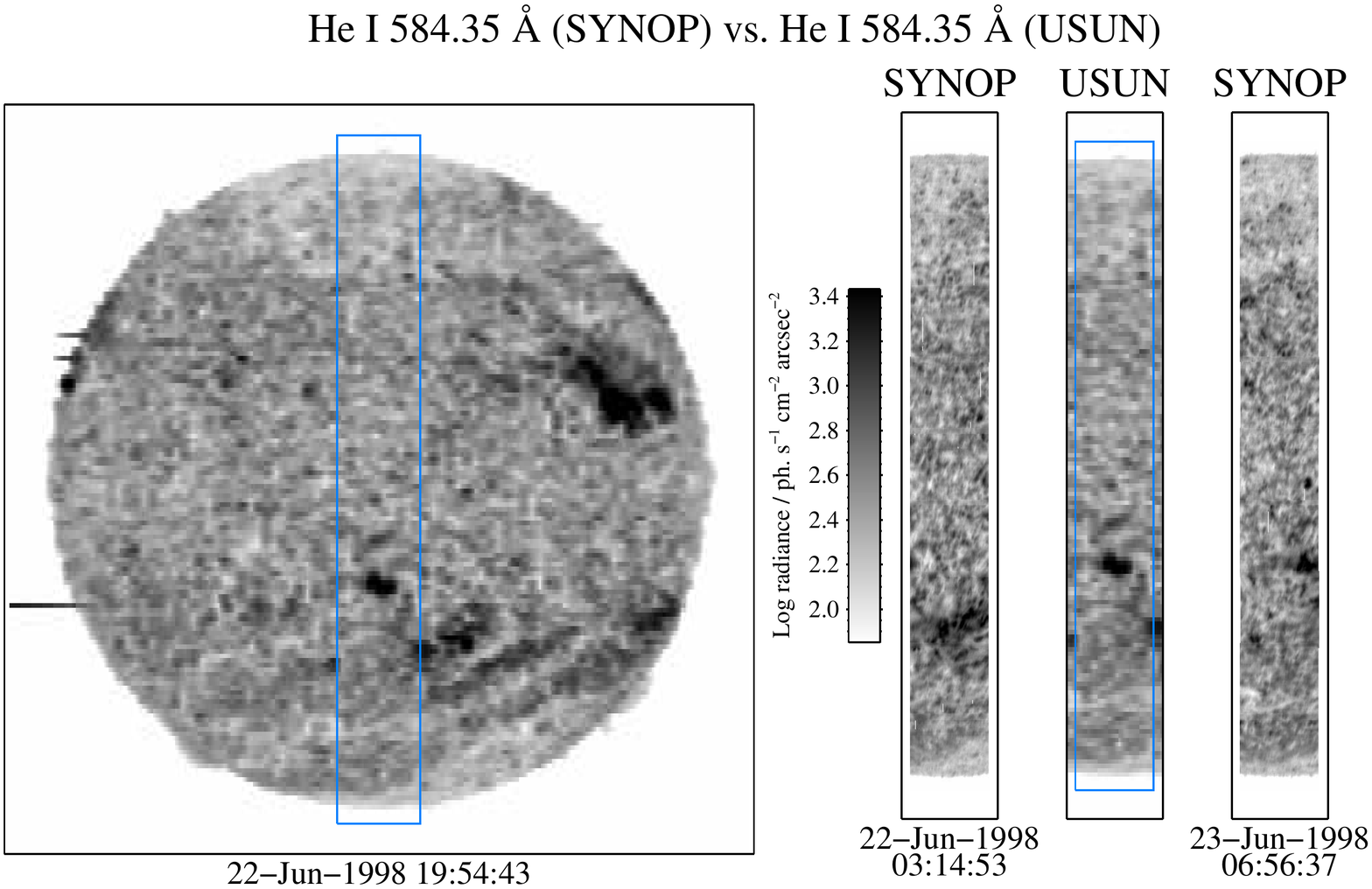}
  \includegraphics[width=0.475\linewidth]%
  {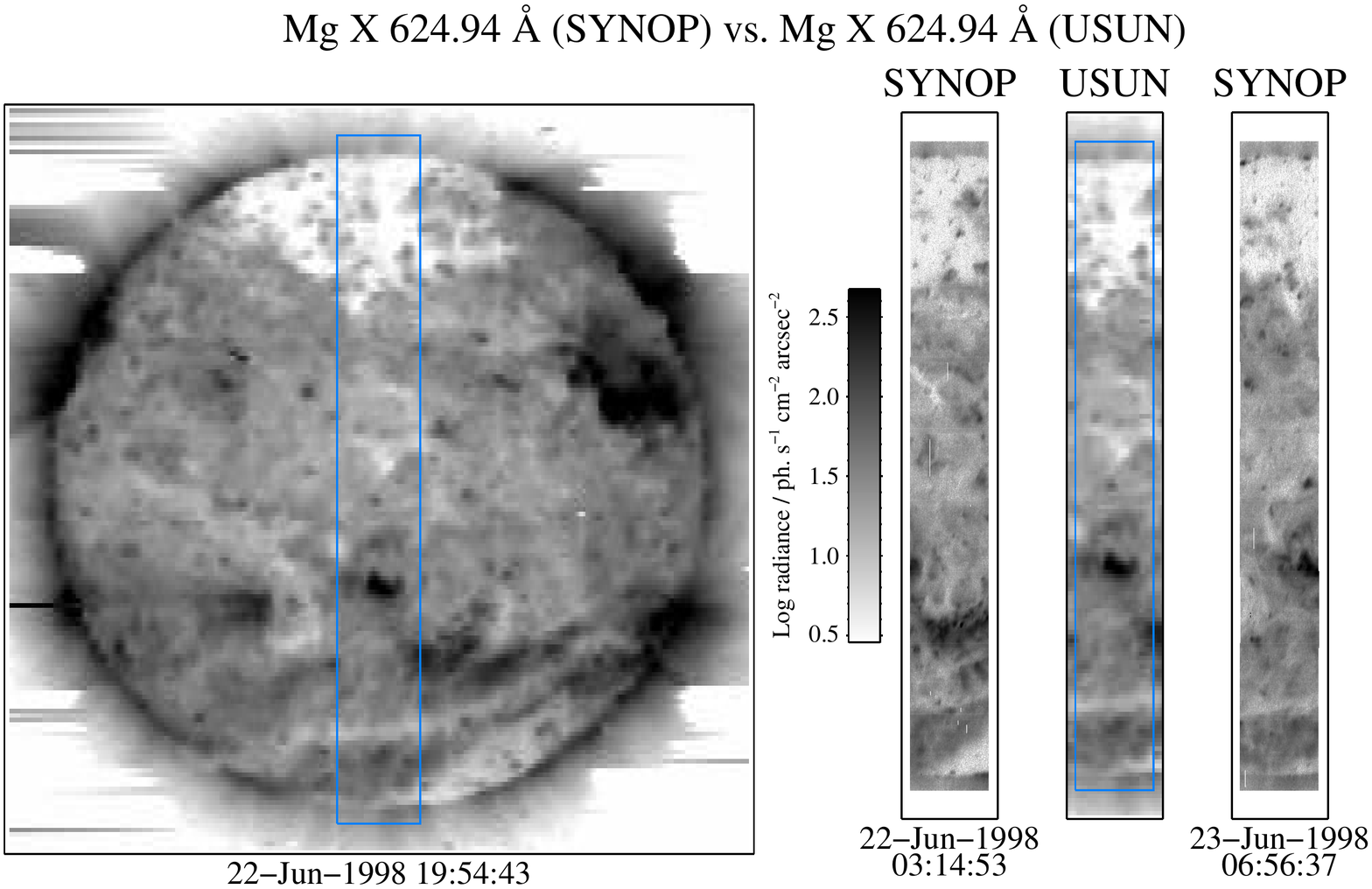}
  \caption{%
    Comparison of USUN and SYNOP data taken on 22 and 23 June 1998. The
    images show the integrated line radiances for the \ion{He}{i} 584~\AA\ and
    \ion{Mg}{x} 625~\AA\ lines in inverted, logarithmic scale
    (brighter areas correspond to fainter line radiances).  
    The boundaries of the SYNOP FOV are also
    indicated.  
  }
  \label{fig:hist_comp_he1_mg10}
\end{figure*}

A comparison of the two data sets is shown in
Fig.~\ref{fig:hist_comp_he1_mg10} using data taken around 22 June 1998. The
USUN mosaic in the \ion{He}{i} 584~\AA\ and \ion{Mg}{x} 625~\AA\ lines,
started on June 22, is compared with the SYNOP scans immediately preceding
and following one day.
The SYNOP FOV (field of view) is indicated by a box in the USUN image.  Apart
from a well developed northern polar coronal hole and from a small, spotless
active region crossing the meridian below the equator, the Sun is largely
quiescent in the SYNOP FOV.  

\begin{figure}[!b]
  \centering
  \includegraphics[clip=true,trim=35 25 50 35,width=\linewidth]{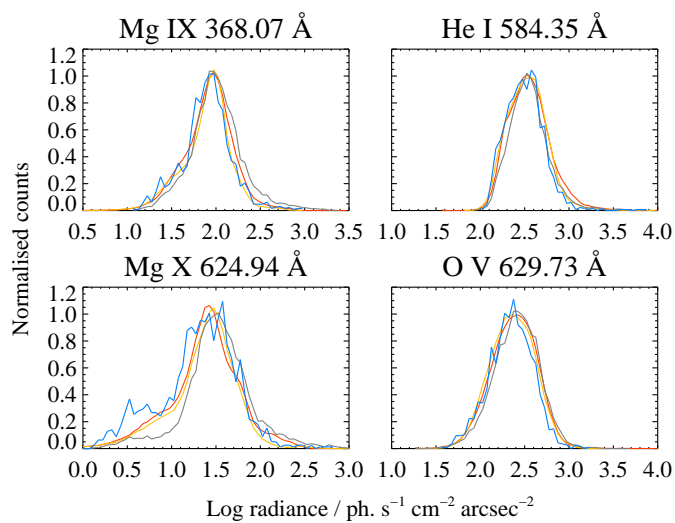}
  \caption{%
    Comparison of line radiance histograms of four representative lines
    (\ion{Mg}{ix} 368~\AA, \ion{He}{i} 584~\AA, \ion{Mg}{x} 625~\AA, and
    \ion{O}{v} 630~\AA) 
    from spectra in the USUN and SYNOP mosaics, as shown in
    Fig.~\ref{fig:hist_comp_he1_mg10}: USUN on-disk data (grey), USUN
    on-disk data within the SYNOP FOV (blue), and data from the two SYNOP
    mosaics (red: mosaic starting on 22 June; orange: mosaic started on 23
    June).
  }
  \label{fig:hist_comp}
\end{figure}

Figure~\ref{fig:hist_comp} shows a comparison of the radiance histograms from
the three data sets, excluding off-limb pixels; in addition, the normalised
histograms obtained from USUN data both for all on-disk spectra and for
spectra within the SYNOP FOV are also shown.  The normalisation factor is the
peak value of each histogram estimated via a Gaussian fit.  Along with the
histograms corresponding to the images of Fig.~\ref{fig:hist_comp_he1_mg10},
we also show the corresponding histograms for two
other strong lines, \ion{Mg}{ix} 368~\AA\ and \ion{O}{v} 630~\AA, in Fig.~\ref{fig:hist_comp}.  In all four
cases, the core of all these histograms is clearly log-normal and corresponds
to the contribution from quiescent areas.  We verified that the same is true
for all the lines in the NIS range for which measurements with reasonable S/N
ratio could be made.

Concerning this property of the quiet-Sun statistical distribution of
radiances, it is useful to recall
early studies of quiet-Sun radiance distribution from the ATM (Apollo telescope
mount) Skylab data
\citep[e.g.:][]{Skumanich-etal:75,Reeves:76}, which typically analysed quiet
Sun histograms as being composed by a Gaussian core and a high-radiance
tail. In some cases a bi-modal distribution was fitted \citep[e.g.:][using CDS
observations]{Gallagher-etal:98} and interpreted in terms of a mixture of supergranular
cell-centre and network contributions.  These interpretations would hint to
different energy dissipation mechanisms between lower and higher radiance
(or low and high magnetic field) regions within the quiescent solar atmosphere.

It was then realised by \cite{Wilhelm-etal:98} and by \cite{Griffiths-etal:99}
using SOHO/SUMER data and by \cite{Pauluhn-etal:00} using both SUMER and CDS
data that a single, log-normal distribution fitted the data better.  A
single-parameter, continuous distribution would thus suggest that the same
heating process is operating both at the edge the network and at the centre of
superganular cells.  Similar properties were inferred for the upper
chromospheric lines Ly-$\alpha$ and Ly-$\beta$ \citep{Curdt-etal:08}.  We
refer to \citet{Fontenla-etal:07} for a more detailed discussion, including a
comparison with distributions of photospheric continuum and magnetic field,
and of \ion{Ca}{ii} K$_3$ intensity.

In addition to a log-normal quiet-Sun contribution, the radiance histograms of
Fig.~\ref{fig:hist_comp} also include 
contributions due to non-quiescent areas.  In
particular, these contributions
are clearly visibile in the hotter lines and correspond
respectively to the coronal hole (\ion{Mg}{x} 625~\AA\ radiances below $\sim
10$ photons s$^{-1}$ cm$^{-2}$ arcsec$^{-2}$) and the small active region
(\ion{Mg}{x} 625~\AA\ radiances above $\sim 100$ photons s$^{-1}$ cm$^{-2}$
arcsec$^{-2}$); an additional contribution comes from near-limb pixels (limb-brightening, more apparent in the histogram from the USUN full FOV in
the figure).  Those contributions are less evident in the histograms of the
other two lines, due to the smaller contrast and/or to the smaller areas of
those features in TR lines but are nevertheless present.  By
analysing the histograms in the corresponding regions of interest, for instance, it can be
shown that the ``hump'' at $\sim 200$ photons s$^{-1}$ cm$^{-2}$ arcsec$^{-2}$
in the \ion{He}{i} histogram does indeed correspond to the contribution of the
northern hole, while the tail above $\sim 10^3$ photons s$^{-1}$ cm$^{-2}$
arcsec$^{-2}$ is the contribution of the active region.

In Sec.~\ref{sec:disc:qs}, we exploit this feature of the EUV radiance
distributions to isolate the quiet-Sun contribution to the radiance histograms
by fitting the main peak of the distribution with a Gaussian.  We thus are 
able to obtain an estimate of the median (central) radiance and width of the quiet Sun
distribution for each SYNOP or USUN study.

In this regard, it is worth mentioning explicitly here that
the mean value of $x$ is
\begin{equation}
  \log <\!x\!> = 
  p+w^2\:(\ln 10)/2 = 
  p+1.15\:w^2 \label{eq:bias} \; ;
\end{equation}
given a normal
distribution of a variable $y$ centred at $p$ and with standard deviation
$w$, and the corresponding log-normal distribution of the variable $x=10^y$.
  The median of $x$, on the other hand, coincides with $10^{p}$.
  Since we analyse the statistical properties of the
  logarithm of radiances in the remainder of this paper, this difference should be kept in mind when
  translating the measured centres of their normal distribution to mean
  radiances.

In any case, a key point is that the statistical distributions of radiances for these lines
are the same in both the SYNOP and USUN data sets:
neither the different spatial resolution nor 
other differences (detector windowing and binning, background level, S/N
ratio, etc.)
produce significant
differences;           the only notable difference in all histograms is that
  USUN histograms restricted to the SYNOP FOV are generally noisier because
  of the lower statistics.     There are some differences in some much weaker
lines (such as the \ion{Fe}{xvi} 360~\AA\ line, also in SYNOP
scans, but hardly detectable in quiet Sun areas) that can be understood in
terms of the effect of the S/N ratio and detector windowing on the
determination of the background continuum.

\subsubsection{Dark halos around active regions}\label{sec:disc:hist:dh}

Because of the presence of various structures on the solar
disk, as noted above, only the core of the radiance histograms can be generally approximated
with a log-normal distribution.

\begin{figure}[!htbp]
  \centering
  \includegraphics[clip=true,trim=540 30 40 50,height=0.90\textheight]{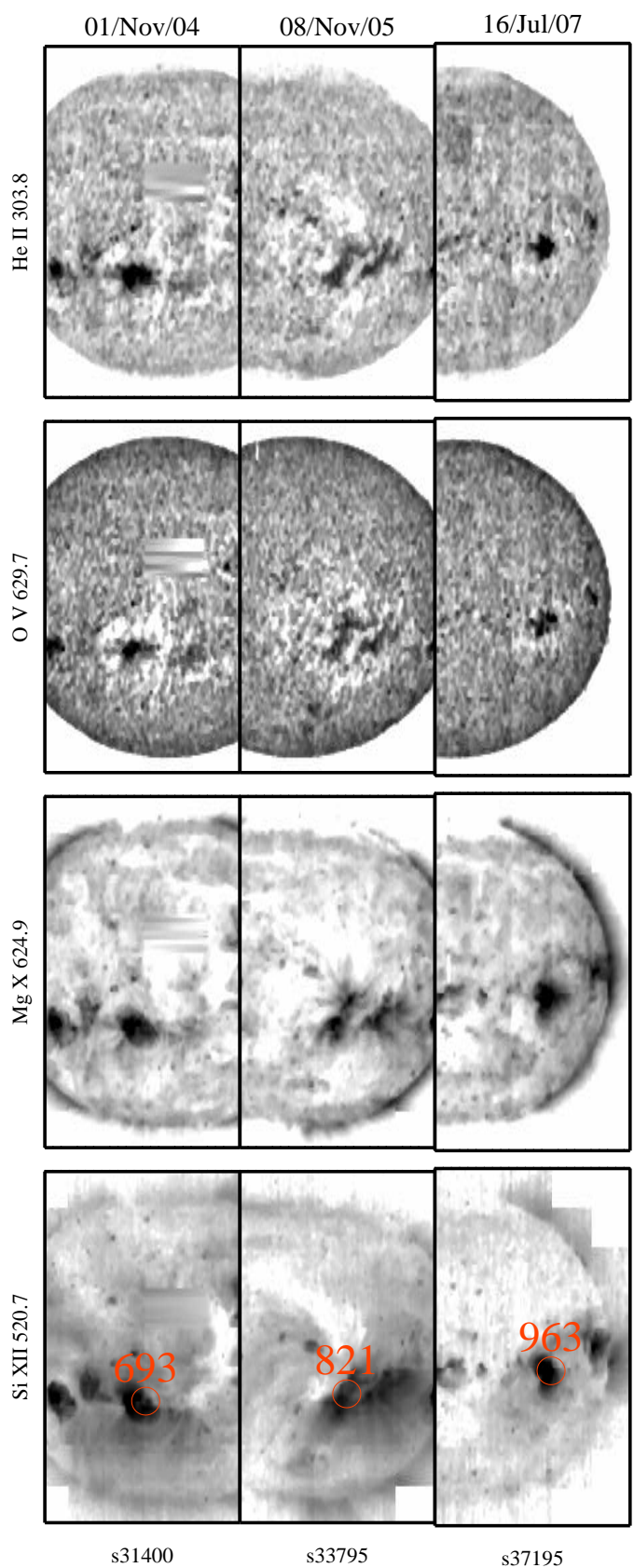}
  \caption{%
      Radiance maps for the lines \ion{He}{ii} 304~\AA, \ion{O}{v} 630~\AA,
      \ion{Mg}{x} 625 \AA, and \ion{Si}{xii} 521 \AA\ (from top to bottom),
      from the CDS USUN mosaics taken on 1 November 2004, 8 November 2005, and 16 July
      2007 (left to right).  The active regions NOAA 693, 821, and 963,
      respectively, are also indicated.  The logarithmic colour display table is inverted:
      higher-radiance areas are darker; for each line, the range of radiances
      is the same for all three dates.  The horizontal stripes near the centre
      of the solar disk in the 2004 mosaics are due to a missing raster, which are
      filled-in here with interpolated values for display purposes.
  }
  \label{fig:dark_halos}
\end{figure}

In the lower end of the tail of the histograms, one of the obvious
contributions comes from coronal holes.  However, a major contribution often
comes from ``halos'' around active regions (Fig.~\ref{fig:dark_halos}) that
sometimes are mistaken as disk coronal holes.  
That figure indeed shows that even around relatively small active regions,
such halos can be seen.  An example is the area south-east of NOAA AR
963 on 16 July 2007.  The polar coronal holes visible on all three dates may
serve as a qualitative indicator of the extent of suppression of line
radiance in those halos.

These structures, which were first identified in chromospheric lines,
have recently been studied by
\cite{DelZanna-etal:11} and \cite{WangYM-etal:11}; the latter also provides a
short review of historical observations.  An earlier review by
\cite{Feldman-etal:00} noted that these regions are not seen in the
\ion{He}{ii} 304~\AA\ line, contrary to genuine coronal holes.  However,
as shown in Fig.~\ref{fig:dark_halos}, these dark canopies, as they are
sometimes called, are easily seen in that line.  The real difference
with coronal holes lies in the observation that these halos are easily seen in
all TR lines, while coronal lines are not always suppressed, perhaps, because
of emission by overarching active region (AR) loops.

Regarding the nature of these regions, \cite{WangYM-etal:11} emphasise their
association with near horizontal fibrils which contain EUV absorbing material
(neutral hydrogen or helium).  However, inspection of images of such regions
taken with SUMER in 1996 \citep{SUMER-Atlas:03} reveals that these regions are seen
as darker areas even in lines at wavelengths longer than 912~\AA; 
a dark halo around an active region can for instance be seen in the \ion{S}{vi}
933~\AA\ image shown in Fig.~1 of \citet{Wilhelm-etal:98}.  Therefore,
the interpretation based on absorption by neutral hydrogen cannot fully
explain those features.  Moreover, as noted by \cite{DelZanna-etal:11}, these
regions are darker even in broadband (thin filters)  Hinode XRT 
X-ray images, where absorption by cool material should be unimportant.
We note that the observations showing that these features are seen in the X-rays, do not mean
that they are hot, because the thin filters receive a significant 
contribution from emission lines formed at and below 1~MK.

The model proposed by \cite{DelZanna-etal:11}, which involves interaction of
active region closed loops with nearby lines opened to the solar wind, could
be more promising, although it still remains to be explained why such regions
are seen in TR lines, while coronal holes are not.  Further discussion on this
issue is in order.  In the context of this paper, it only needs to be remarked
that, indeed, these structures are common and seen in spectral features
that form on a wide range of temperatures from the chromosphere through the
transition region up to lines forming around $\sim 10^6$~K.

\subsection{Some properties of quiet Sun regions}\label{sec:disc:qs}

We now turn our attention to the core of radiance histograms, or the quiet
Sun, thus filtering out the effects by active regions (ARs), coronal holes
(CHs), and dark halos (DHs) to analyse the variability of their properties
during the activity cycle.  Such a study, however, may be hampered by
non-negligible centre-to-limb variations. We therefore first examine the
variation across the solar disk of the median radiance
(Sec.~\ref{sec:disc:qs:c2l}), and then we discuss the mean properties of the
radiance histograms (Sec.\ref{sec:disc:qs:cycle}).

\subsubsection{Centre-to-limb variation in the quiet Sun radiances}\label{sec:disc:qs:c2l}

\begin{figure*}[!t]
  \centering
  \includegraphics[clip=true,trim=-50 90 0 120,width=0.95\linewidth]{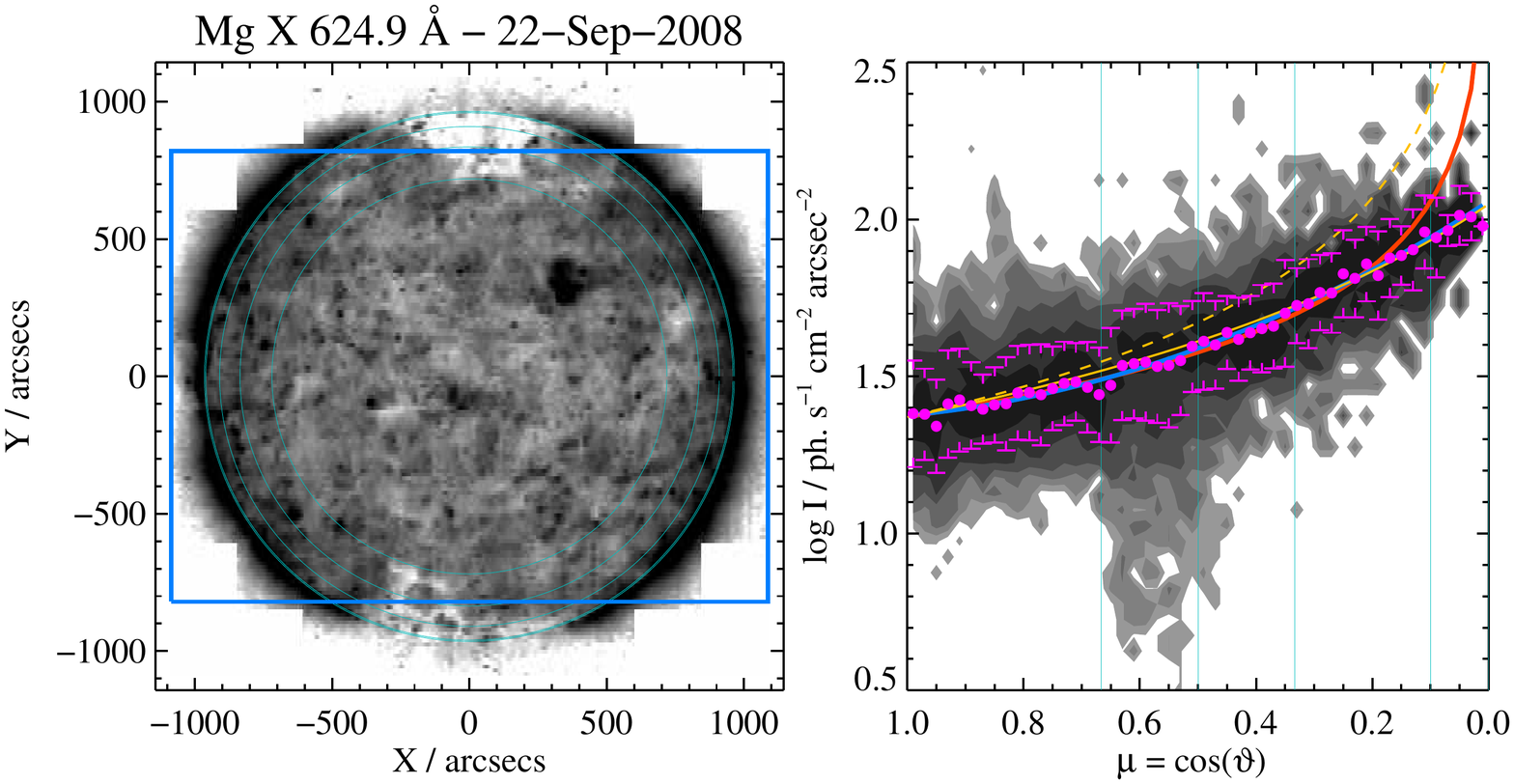}\\
  \includegraphics[clip=true,trim=-59 90 0 120,width=0.95\linewidth]{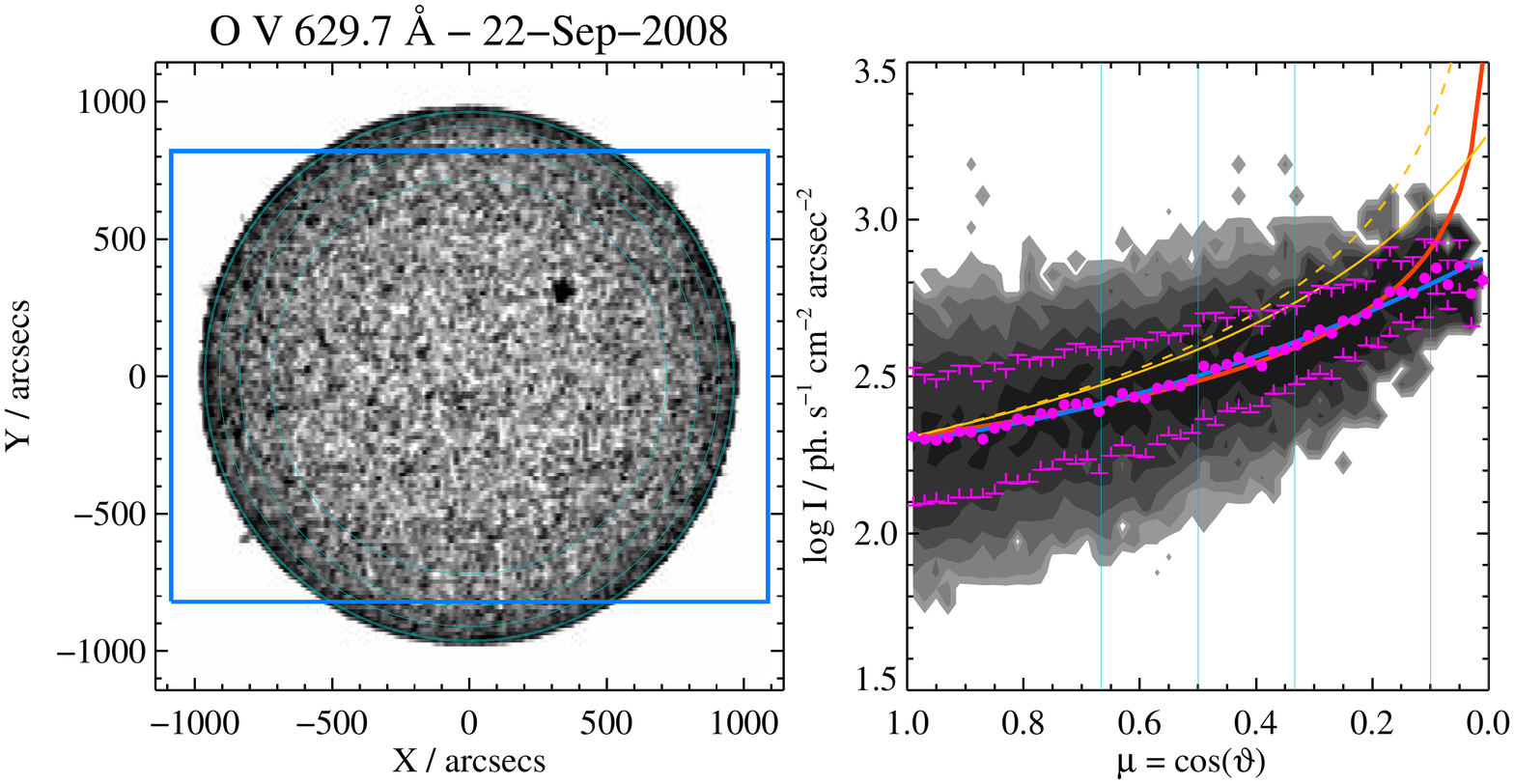}
  \caption{%
    Maps of \ion{Mg}{x} 625~\AA\           and \ion{O}{v} 630~\AA\ radiances (left-hand panels) in a logarithmic,
    inverted scale (brighter areas correspond to fainter line radiances) from
    a USUN mosaic obtained on September 2008.
    The corresponding bi-dimensional histograms vs.\ $\mu$ 
    are shown in the upper-right and lower-right panel, respectively;  
    the position and width of the histogram peaks are also shown
    (magenta dots and tee-shaped symbols). 
    The power-law and quadratic curves fitting the peaks of the
    $\log I$ vs.\ $\mu$ histograms are shown as red and blue lines,
    respectively.  
    Curves derived from plane-parallel and spherical optically thin
    models are shown as dashed and continuous orange lines, respectively, in
    both right-hand panels.
    Distances corresponding to $\mu = 0$ (white light limb), $1/10$, $1/3$, $1/2$, and
    $2/3$ are shown as circles in the left-hand panel and as vertical lines
    in the right-hand panels.  
    The rectangle corresponding to the FOV limited
    to $|Y|<0.85\: R_\odot$, which is used to compute the histograms vs.\ $\mu$, is
    also shown.
  }
  \label{fig:c2l_images}
\end{figure*}

The centre-to-limb variation in line radiance is interesting \textit{per se},
allowing for instance distinguishing between optically thin and thick lines
but is also relevant for estimates of the disk-integrated line emission,
  which is proportional to the line irradiance in turn, aside from a constant
  and neglecting off-limb emission \citepalias[a good approximation for all
  but the hottest lines, as shown in][]{DelZanna-Andretta:11}.

          For an empirical estimate of the centre-to-limb variation in the
  quiet Sun component, we 
employed the same technique described in \cite{Andretta-etal:03}:
for each annulus of radial distance $r$ from disk centre and width $\Delta
r=0.02\:R_\odot$, we computed the histogram of radiances, thus obtaining a
bi-dimensional histogram as function of $\rho=r/R_\odot$.  To characterise the
dependence of radiance from radial distance on the solar disk, it is usually
convenient to consider the centre-to-limb variations as function of $\mu$, the
cosine of the heliocentric angle $\theta$.  We also computed, therefore,
bi-dimensional histograms as a function of $\mu$, adopting a bin size
$\Delta\mu=0.02$.   

For selected lines in the CDS/NIS range and for each USUN or SYNOP mosaic
analysed, we computed a similar bi-dimensional histogram.  For the
calculation of the $I$ vs.\ $\mu$ histogram in USUN mosaics, we excluded the
polar regions to minimize
the contribution of polar coronal holes during the minimum of solar activity, 
since we are mostly
interested in the properties of the gross QS centre-to-limb variation.
Adopting a rectangular coordinate system, $(X,Y)$, with its origin at the
centre of the solar disk and with the $Y$ axis oriented along the central
meridian, we then selected pixels in the  region determined by $|Y|<0.85\: R_\odot$.

Figure~\ref{fig:c2l_images} shows two examples of these bi-dimensional histograms
obtained from a USUN mosaic taken during the latest solar minimum for the
coronal line %
\ion{Mg}{x} 625~\AA\ %
and the TR line \ion{O}{v} 630~\AA. %
The line-radiance maps are shown in the left-hand panels with
circles indicating some representative heliocentric distances. The corresponding
bi-dimensional histogram as function of 
$\mu$, computed from
on-disk spectra within the blue rectangle, are shown in the right-hand panels.

The measured positions of the peaks of the radiance histograms as function of
$\mu$ are shown in Fig.~\ref{fig:c2l_images} with the estimated
widths (standard deviation) of the distributions.  Note that the tails of the
histograms are due to the small active region at $\rho\sim 0.5$ ($\mu\sim
0.85$) and to the polar coronal holes ($\rho>0.75$, $\mu<0.67$).  

In a \emph{plane-parallel} geometry, the total radiance of optically thin
lines varies as $\sim \mu^{-1}$.
This function is shown in Fig.~\ref{fig:c2l_images} as a dashed orange line.
It is sometimes assumed 
\citep[e.g.:][]{Wilhelm-etal:98} that solar optically thin lines should
exhibit a $\sim \mu^{-1}$ limb brightening, while significant departures from
such a dependence would be characteristic of optically thick lines or of
lines affected (below the Lyman edge) by absorption due to cool material.
However, as shown by \cite{Andretta-etal:03}, a
\emph{spherical, exponentially stratified} model of line emissivity can fit
observed centre-to-limb behaviour of coronal lines without invoking
optical thickness effects.  The centre-to-limb behaviour from such a model,
which adopts a scale height corresponding to the nominal formation temperature
of the Mg$^{+9}$ ion ($\log T_\mathrm{max} = 6.0$) (shown in Fig.~\ref{fig:c2l_images} as a
continuous orange line) fits well to the observed average
centre-to-limb variation of line radiances.  
  The match, while still better than the function $\mu^{-1}$, is less good in
  the case of the O$^{+4}$ ion ($\log T_\mathrm{max} = 5.4$).  It, however,
  still serves as a warning against interpreting the empirical centre-to-limb
  variation in that line only in terms of optical thickness effects.

  For simplicity, we fitted the logarithm of these peak radiances as function
  of $\mu$ with a quadratic function:
  \begin{equation} \log I(\mu) = \log I^\mathrm{q}(1)
    +a\:(1-\mu)+b\:(1-\mu)^2.  \label{eq:c2l_pol}%
\end{equation}
To
facilitate comparisons with the results of \cite{Pietarila-Judge:04}, we
also computed power-law fits: 
\begin{equation}
I(\mu) = I^\mathrm{p}(1)\; \mu^\alpha\; ; \label{eq:c2l_pow}
\end{equation}
note that the
value $\alpha=-1$ corresponds to the plane-parallel finite-slab,
optically-thin case.  
  In the fit procedure, we only considered histogram bins corresponding to
  $\mu>0.1$, which are further away than $\sim 5\arcsec$ from the nominal,
  white-light limb.  %
The resulting fitting curves are shown in the right-hand panels of
Fig.~\ref{fig:c2l_images} as a blue and red curve, respectively.

          Once the average centre-to-limb variation in a line has been
  determined, it is then possible to derive its mean disk radiance (and thus
  its irradiance) from a measurement at a given $\mu$ on the solar disk,
  through the multiplicative factor $f_\mathrm{cl}(1)$, where
  $f_\mathrm{cl}(\mu)$ is defined by \cite{Andretta-etal:03} as follows:  
\begin{equation}
  f_\mathrm{cl}(\mu) \equiv 2\:\int_0^1
  \mu'\,\frac{I(\mu')}{I(\mu)}\,\mathrm{d}\mu'\; . \label{eq:fcl}
\end{equation}
The above correction factor computed at $\mu=1$ is normally the most
useful value in the case of observations near disk-centre of the Sun in a quiescent
state: in the remainder, we refer to that value as simply $f_\mathrm{cl}$.  In
the case of Eq.~\ref{eq:c2l_pol} and Eq.~\ref{eq:c2l_pow}, the correction
factors $f^q_\mathrm{cl}$ and $f^p_\mathrm{cl}$, respectively, can both be
computed analytically and will be discussed later.

  The centre-to-limb variation measured on each USUN or SYNOP mosaic have also
  been averaged to produce mean empirical centre-to-limb variations.  The
  averages were done separately for dates corresponding to solar minimum and
  solar maximum.
More specifically, we considered the
following time intervals for the minima and maximum of solar activity:
\begin{description}
\item[1996-1997 minimum:] 
  SYNOP mosaics between 2 April 1996 and 17 May 1997 (no USUN studies in
  that time interval);
\item[2007-2009 minimum:] 
  USUN mosaics between 29 October 2007 and 30 November 2009;
  SYNOP mosaics beween 26 November 2007 and 1 December 2009;
\item[Maximum:] 
  USUN mosaics between 13 March 2000 and 6 August 2002;
  SYNOP mosaics between 5 June 2000 and 20 October 2002.
\end{description}

\begin{figure*}[!hbt]
  \centering
  \includegraphics[clip=true,trim=-10 30 0 55,width=\linewidth]{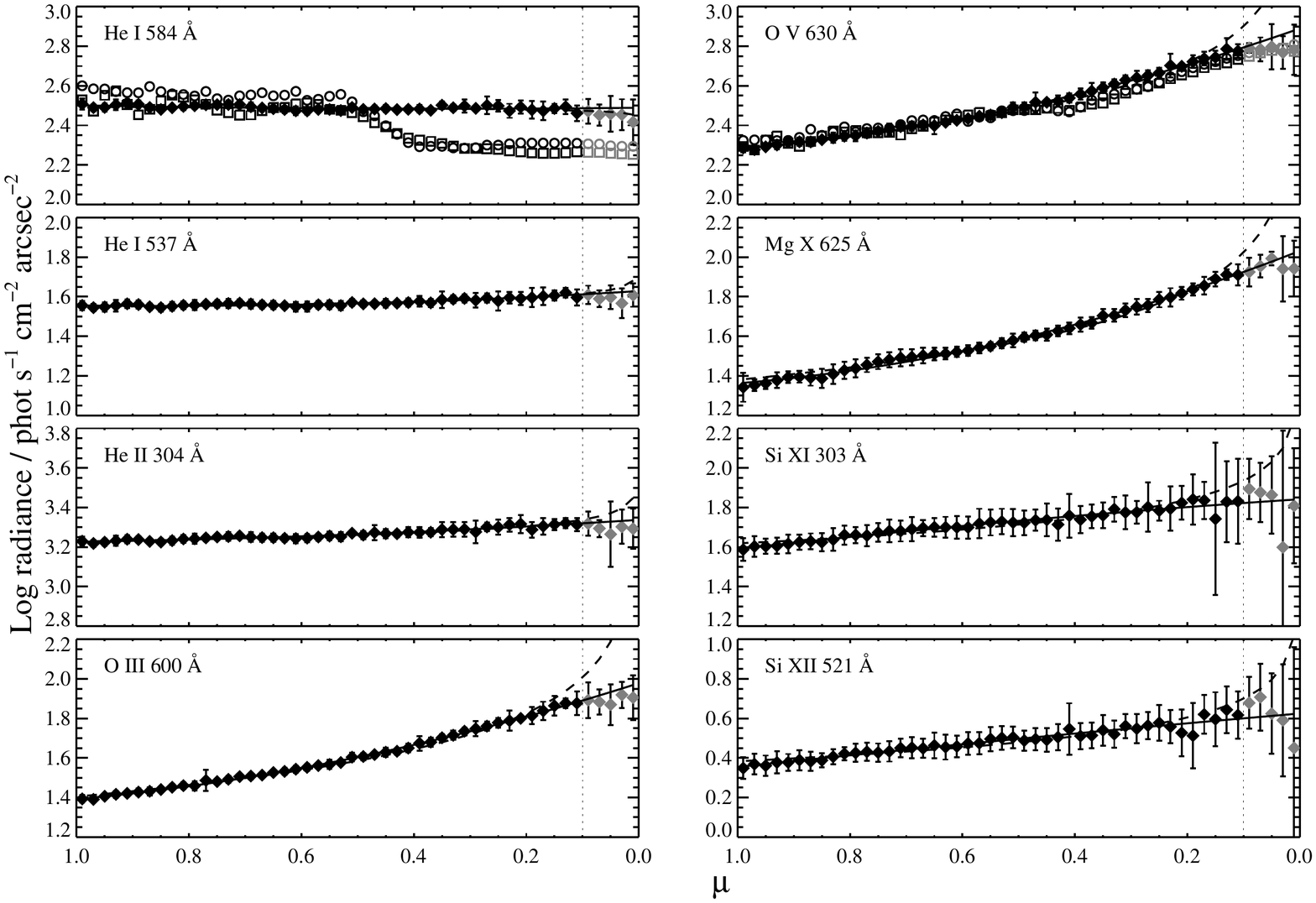}
  \caption{%
      Average centre-to-limb variation in several strong lines from USUN
      mosaics (diamonds). 
      For the \ion{He}{i} 584~\AA\ and \ion{O}{v} 630~\AA,
      where SYNOP data are available, squares indicate SYNOP data during
      the 2007-2009 solar minimum, while circles denote data during the
      1996-1997 minimum. 
      The quadratic (solid lines) and power-law (dashed lines) fits to the
      USUN data are also shown. Grey points ($\mu<0.1$)
      indicate data not used in the actual fits. %
  }
  \label{fig:c2l_profs}
\end{figure*}

  The average, empirical profiles for the solar minimum are shown in
  Fig.~\ref{fig:c2l_profs}.  At each value of $\mu$, the standard deviation of
  the peak radiances for all the solar minimum dates considered 
  provides an estimate for the errors.  The average profiles at
  solar maximum are to be considered less reliable, given the noise introduced
  by solar non-quiescent structures, and are therefore not shown in that
  figure for clarity, but they overlap well with the more accurate solar
  minimum centre-to-limb profiles.

  The mean empirical profiles are then again fitted with the same functions
  described above.  The resulting fitting function are also shown in
  Fig.~\ref{fig:c2l_profs} in the case of solar minimum, and the fit
  coefficients with their standard deviations are listed in
  Table~\ref{tab:c2l_fits} for both activity minima and maximum.

The \ion{He}{i} 584~\AA\ and \ion{O}{v} 630~\AA\ lines have been included in
the definition of SYNOP mosaics since the start of the SOHO mission; however,
we do not list the results for the \ion{He}{i} line from SYNOP mosaics here,
because they are normally affected by the presence of polar coronal holes,
especially during the minimum of solar activity. %
The effect of coronal holes is indeed clearly seen in the SYNOP average
profiles of Fig.~\ref{fig:c2l_profs}.

\begin{figure}[!b]
  \centering
  \includegraphics[clip=true,trim=20 40 40 40,width=\linewidth]{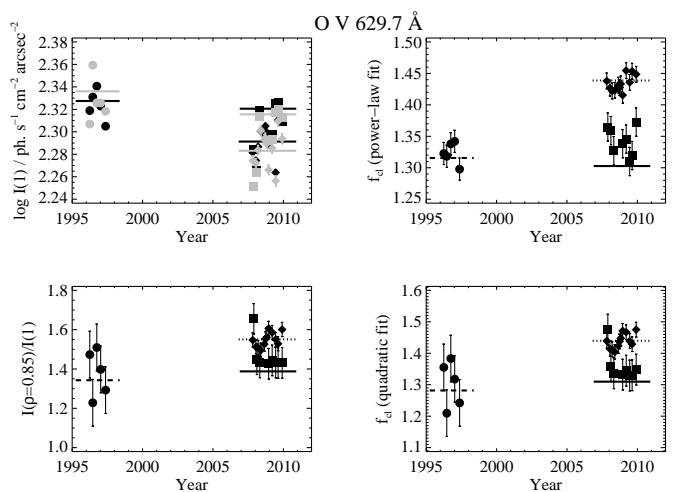}
  \caption{%
    Parameters characterising 
    the centre-to-limb variation in the \ion{O}{v} 630~\AA\
    line during the last two solar minima.
    Diamonds: USUN; squares: SYNOP (2007-2009), circles: SYNOP (1996-1997).
    The average values are indicated by
    horizontal lines: dotted (USUN), solid (SYNOP, 2007-2009), and dashed
    (SYNOP, 1996-1997). 
    In the upper-left panel, the centre-disk intensity, $I(1)$, from the
    power-law fit are in black, while
    results from quadratic fits are in grey.  
      The correction factors $f_\mathrm{cl}$, whose average values are
      listed also in
      Table~\ref{tab:c2l_fits}, are shown in the right-hand panels; the value
      of centre-to-limb function at $\rho=0.85$ (i.e. $\mu\sim 0.52$),
      derived from the quadratic fit of Eq.~\ref{eq:c2l_pol}, is shown in the
      lower-left panel. %
  }
  \label{fig:c2l_fits}
\end{figure}

Coronal holes, on the other hand, are almost undetectable in \ion{O}{v} 630
\AA\ images, although a hint of a slight decrease of radiances can
  be seen in Fig.~\ref{fig:c2l_profs} in the SYNOP average profiles below
  $\mu\sim 0.4$.  We therefore use SYNOP data to measure the
centre-to-limb variation in that line along the full solar cycle from the
1996--1997 to the 2007--2009 minima.  
  Figure~\ref{fig:c2l_fits} shows the variability of the centre-to-limb fitted
  function for the \ion{O}{v} 630 \AA\ line in mosaics taken during the
  minimum of solar activity from both the mean profile and the
  individual dates.  Rather than showing the fit parameters $\alpha$ or $a$
  and $b$ listed in Table~\ref{tab:c2l_fits}, we have chosen to show the
  variation in the centre-to-limb function computed at $\rho=0.85$ and the
  factor $f_\mathrm{cl}$, which are quantities that are more immediately illustrative of the
  properties of the centre-to-limb variation.
No significant variation in the
centre-to-limb variation is apparent in these data.  The centre-disk radiance
in our data,
$\log I(1)$, is constant by definition, because the long-term correction to
the CDS/NIS radiometric calibration worked out in
\citetalias{DelZanna-etal:10} assumes a constant mean QS radiance with time.
On the other hand, the variation in the radiance from centre to
  limb
does not
depend on the radiometric calibration.  Therefore, its
invariance between the two solar minima suggests that the structure
of the transition region did not change.

  Note also that parameters of the centre-to-limb functions given in
  Table~\ref{tab:c2l_fits} do not change between solar miminum and maximum
  (except for the hotter lines).  The invariance of the centre-to-limb
  function is best seen by examining the factors $f_\mathrm{cl}(1)$, which 
  are consistent with constant values in all cases within the stated
  errors, usually within one or two standard deviations.

It is useful to note also that the parameters obtained from USUN and SYNOP mosaics are consistent
within the errors despite the different FOVs and spatial
resolutions.  However, \cite{Wilhelm-etal:98} noticed that the limb
brightening of TR lines above coronal holes is different compared to
the QS (a property of TR limb brightenings already described earlier by, e.g.,
\citealt{Doschek-etal:76,Feldman-etal:76}).  To verify whether this effect could
account for the residual differences between SYNOP and USUN results shown in Fig.~\ref{fig:c2l_fits} and listed in
Table~\ref{tab:c2l_fits}, we computed the centre-to-limb variation in USUN
mosaics as explained above but in only in a strip along the central meridian
similar to the SYNOP FOV (cf. 
Sec.~\ref{sec:disc:hist:comp}).

  For the \ion{O}{v} 630~\AA\ line, the factor $f^q_\mathrm{cl}$ during the
  2007-2009 minimum, for instance, changes from $1.44\pm 0.02$ to $1.27\pm
  0.06$ when restricting the analysis to the SYNOP FOV, which is in excellent agreement with the value obtained from SYNOP data.  A
  similar change is less clearly seen in the fainter and cooler \ion{O}{iii}
  600~\AA: the factor changes from $1.45\pm 0.02$ to $1.34\pm 0.11$, while the
  statistics in the \ion{O}{iv} 554~\AA\ multiplet gives even larger
  uncertainties, thus hiding this effect.
While we can state that our results are
consistent with the results of \cite{Wilhelm-etal:98} concerning the effects
of coronal holes in TR lines, the worse statistics of USUN mosaics (due to fewer spectra in
the central meridian strip)
does not permit an accurate measurement of this effect.  

  An inspection of the power-law exponents listed in Table~\ref{tab:c2l_fits}
  shows that in all cases $\alpha$ is negative with the exception perhaps of the \ion{He}{i}
  584~\AA\ line, indicating a limb brightening, and its
  absolute value is always significantly smaller than one.  We once again stress that
  departures from the value $\alpha=-1$ do not imply significant line optical
  thicknesses but only an inadequacy of the plane-parallel finite-slab,
  optically-thin model.  The majority of lines in the CDS wavelength range can
  still be assumed to be (effectively) optically thin, since their limb
  brightening can be modelled (very well in many cases) by an exponentially
  stratified emissivity in a spherical geometry, as shown in
  Fig.~\ref{fig:c2l_images}. This takes into account that the quiet-Sun component
  in the hotter lines, such as \ion{Si}{xii} 521~\AA\ line, is less clearly
  defined, and thus their limb brightening fits are more uncertain.

  The only likely exceptions are the \ion{He}{i} and \ion{He}{ii} lines, which
  all exhibit very shallow centre-to-limb variations.  While neither a
  plane-parallel slab nor a spherical shell geometrical model are likely to be
  reasonable representations of the chromosphere or of the lower TR, it is
  also likely that a significant optically thickness in those lines needs to
  be invoked, as already pointed out by \cite{Pietarila-Judge:04}. The
  values of $\alpha$ computed by the latter authors, however, are still significantly larger in
  absolute value than the observed values we present
  here. That is, they derive a limb brightnening in the helium resonance lines from their
  calculations which is still too strong.  %
    It is useful to note that the centre-to-limb variation in \ion{H}{i}
    Ly-$\alpha$, as measured with SUMER by
    \citet{Curdt-etal:08} is also flat.

For what concerns a comparison to other measurements, we mention the work
by \cite{Mango-etal:78} here, who fitted measurements for the \ion{He}{i} 584~\AA,
\ion{He}{i} 537~\AA, and \ion{He}{ii} 304~\AA\ lines with a linear function,
$I(\mu)=I(1)\:[1+C\:(1-\mu)]$,
with the coefficients $C=0.10$, $C=0.47$, and $C=0.31$, respectively. 
  Note that the coefficient $C$ to a second order can be related to
  coefficients $a$ and $b$ of Table~\ref{tab:c2l_fits}, as $a\simeq
  C/\ln 10$, and $b\simeq -C/(2\ln 10)$.
Those fits were however made from relatively few data points and on a narrow
range of $\mu$: the data for the \ion{He}{i} 537~\AA\ line, in particular,
were derived from the interval $0.2<\mu<0.6$.
We thus do not regard a comparison with the results of our fits as very
significant. 

The factor $f_\mathrm{cl}$ can be compared with the factor
$\bar{L}_\lambda/L_\lambda(0)$ that are given in Table~4 of \cite{Wilhelm-etal:98},
although the latter is obtained from an actual integration of data from
full-disk scans, not from an average centre-to-limb variation.  The only lines
in common between Table~4 of \cite{Wilhelm-etal:98} and
Table~\ref{tab:c2l_fits} are the \ion{He}{i} 584~\AA\ and \ion{O}{v} 630~\AA\
lines: the factor given for the former line is
$\bar{L}_\lambda/L_\lambda(0)=1.01$, which is consistent with the values of the order
of unity we have found. For the latter line, the value $1.59$ is
higher than the value of $\sim 1.3$ from the QS centre-to-limb variation
that is measured in the present work.  The value given by \cite{Wilhelm-etal:98} for
the \ion{O}{iv} 788~\AA\ line, $\bar{L}_\lambda/L_\lambda(0)=1.46$ is similar
to the value we find for the \ion{O}{iv} 554~\AA\ multiplet,
$f_\mathrm{cl}\sim 1.4$.

          It should be noted that  these factors allow an estimate of the disk irradiance due to QS
emission only (no active regions nor coronal holes are accounted for).
Furthermore, as discussed in \citetalias{DelZanna-etal:10}, it is
also necessary to account for off-limb emission for most lines
with formation temperature of the order of $\sim 1$~MK or higher.

For the remainder of this work in any case, we adopt the quadratic fits (Eq.~\ref{eq:c2l_pol}) obtained
from USUN mosaics during the 2007--2009 minimum of solar activity as a reference
centre-to-limb variation.

\begin{sidewaystable*}[!hbt]
  \tiny
  \centering
  \begin{tabular}[tb]{lr@{\,}lr@{\,}lr@{\,}lr@{\,}lr@{\,}lr@{\,}lr@{\,}l}
    \hline
    \hline
    Data set & 
    \multicolumn{6}{c}{power law} &
    \multicolumn{8}{c}{quadratic} \\
             &
    \multicolumn{2}{c}{$\log I^\mathrm{p}(1)$}    & 
    \multicolumn{2}{c}{$\alpha$} &
    \multicolumn{2}{c}{$f^\mathrm{p}_\mathrm{cl}$} &
    \multicolumn{2}{c}{$\log I^\mathrm{q}(1)$}    & 
    \multicolumn{2}{c}{$a$} &
    \multicolumn{2}{c}{$b$} &
    \multicolumn{2}{c}{$f^\mathrm{q}_\mathrm{cl}$} \\
    \hline
 \multicolumn{15}{l}{\ion{He}{i} 584 \AA:}\\
USUN[2007-2009] : &      2.494&(0.0045)&       0.020&(0.014)&       0.9902&(0.0095)&       2.502&(0.0088)&      -0.055&(0.048)&         0.04&(0.057)&        0.975&(0.025)\\ 
USUN[maximum]   : &      2.605&(0.012)&         0.03&(0.035)&        0.987&(0.023)&         2.71&(0.028)&        -0.54&(0.13)&          0.54&(0.14)&          0.81&(0.096)\\ 
SYNOP[1996-1997]: &      2.590&(0.0067)&       0.379&(0.016)&        0.841&(0.011)&        2.597&(0.012)&        -0.13&(0.073)&        -0.28&(0.078)&        0.827&(0.066)\\ 
SYNOP[2007-2009]: &      2.538&(0.0072)&       0.331&(0.013)&        0.858&(0.025)&        2.546&(0.013)&        -0.12&(0.075)&        -0.25&(0.076)&        0.840&(0.030)\\ 
SYNOP[maximum]  : &      2.535&(0.021)&         0.12&(0.046)&        0.943&(0.080)&         2.71&(0.046)&        -0.83&(0.20)&          0.67&(0.20)&          0.69&(0.12)\\  
 \multicolumn{15}{l}{\ion{He}{i} 537 \AA:}\\
USUN[2007-2009] : &      1.548&(0.0051)&      -0.067&(0.015)&        1.034&(0.0095)&       1.556&(0.0095)&       -0.03&(0.051)&         0.10&(0.059)&         1.02&(0.026)\\ 
USUN[maximum]   : &      1.645&(0.011)&        -0.02&(0.028)&         1.01&(0.027)&         1.73&(0.026)&        -0.43&(0.12)&          0.46&(0.12)&          0.86&(0.11)\\  
 \multicolumn{15}{l}{\ion{He}{i} 522 \AA:}\\
USUN[2007-2009] : &     0.9754&(0.0053)&      -0.141&(0.014)&        1.076&(0.0093)&       0.984&(0.0099)&       -0.02&(0.052)&         0.16&(0.057)&         1.05&(0.028)\\ 
USUN[maximum]   : &       1.04&(0.011)&        -0.11&(0.030)&         1.06&(0.022)&         1.13&(0.025)&        -0.43&(0.12)&          0.55&(0.12)&          0.89&(0.11)\\  
 \multicolumn{15}{l}{\ion{He}{i} 516 \AA:}\\
USUN[2007-2009] : &      0.745&(0.0086)&       -0.10&(0.022)&        1.055&(0.010)&        0.752&(0.017)&        -0.02&(0.086)&         0.12&(0.094)&         1.04&(0.028)\\ 
USUN[maximum]   : &      0.863&(0.011)&        -0.11&(0.028)&         1.06&(0.031)&        0.974&(0.025)&        -0.53&(0.12)&          0.64&(0.12)&          0.86&(0.11)\\  
 \multicolumn{15}{l}{\ion{He}{ii} 304 \AA:}\\
USUN[2007-2009] : &      3.228&(0.0048)&       -0.11&(0.015)&         1.06&(0.014)&        3.227&(0.0087)&        0.04&(0.049)&        0.070&(0.057)&         1.06&(0.030)\\ 
USUN[maximum]   : &      3.300&(0.011)&       -0.050&(0.032)&         1.03&(0.034)&        3.357&(0.023)&        -0.29&(0.11)&          0.37&(0.12)&         0.921&(0.084)\\ 
 \multicolumn{15}{l}{\ion{O}{iii} 600 \AA:}\\
USUN[2007-2009] : &      1.404&(0.0039)&      -0.603&(0.012)&        1.431&(0.013)&        1.391&(0.0065)&        0.26&(0.037)&         0.32&(0.044)&         1.45&(0.018)\\ 
USUN[maximum]   : &      1.444&(0.0077)&      -0.553&(0.023)&         1.38&(0.023)&         1.44&(0.016)&         0.15&(0.080)&         0.41&(0.088)&         1.37&(0.038)\\ 
 \multicolumn{15}{l}{\ion{O}{iv} 554 \AA:}\\
USUN[2007-2009] : &      2.081&(0.0059)&      -0.675&(0.020)&         1.51&(0.023)&        2.072&(0.0097)&        0.26&(0.057)&         0.39&(0.070)&         1.50&(0.038)\\ 
USUN[maximum]   : &      2.080&(0.0084)&      -0.597&(0.026)&         1.43&(0.040)&        2.087&(0.017)&         0.13&(0.089)&         0.46&(0.10)&          1.38&(0.046)\\ 
 \multicolumn{15}{l}{\ion{O}{v} 630 \AA:}\\
USUN[2007-2009] : &      2.291&(0.0047)&      -0.610&(0.015)&        1.439&(0.013)&        2.283&(0.0079)&        0.22&(0.044)&         0.39&(0.052)&         1.44&(0.023)\\ 
USUN[maximum]   : &      2.342&(0.0088)&      -0.513&(0.023)&         1.34&(0.020)&        2.346&(0.019)&        0.088&(0.088)&         0.45&(0.093)&         1.32&(0.047)\\ 
SYNOP[1996-1997]: &      2.327&(0.0063)&      -0.480&(0.014)&         1.32&(0.018)&        2.336&(0.010)&         0.05&(0.062)&         0.46&(0.067)&         1.28&(0.074)\\ 
SYNOP[2007-2009]: &      2.321&(0.0071)&      -0.465&(0.015)&         1.30&(0.022)&        2.316&(0.013)&         0.12&(0.064)&         0.39&(0.066)&         1.31&(0.049)\\ 
SYNOP[maximum]  : &      2.312&(0.012)&       -0.541&(0.025)&         1.37&(0.055)&         2.36&(0.024)&        -0.20&(0.12)&          0.77&(0.13)&          1.20&(0.11)\\  
 \multicolumn{15}{l}{\ion{Mg}{x} 625 \AA:}\\
USUN[2007-2009] : &      1.380&(0.0083)&      -0.644&(0.021)&         1.48&(0.032)&         1.36&(0.016)&         0.23&(0.075)&         0.44&(0.081)&         1.50&(0.089)\\ 
USUN[maximum]   : &       1.59&(0.026)&        -0.61&(0.067)&          1.4&(0.23)&          1.67&(0.055)&        -0.29&(0.26)&          0.93&(0.28)&           1.2&(0.41)\\  
 \multicolumn{15}{l}{\ion{Si}{xi} 303 \AA:}\\
USUN[2007-2009] : &      1.619&(0.013)&        -0.32&(0.046)&         1.19&(0.065)&         1.59&(0.022)&         0.32&(0.14)&         -0.07&(0.18)&           1.3&(0.13)\\  
USUN[maximum]   : &       2.21&(0.026)&        -0.44&(0.094)&          1.3&(0.26)&          2.35&(0.058)&        -0.59&(0.28)&           1.1&(0.32)&           1.0&(0.43)\\  
 \multicolumn{15}{l}{\ion{Si}{xii} 521 \AA:}\\
USUN[2007-2009] : &      0.383&(0.013)&        -0.32&(0.045)&          1.2&(0.12)&         0.354&(0.024)&         0.30&(0.14)&         -0.03&(0.17)&           1.3&(0.17)\\  
USUN[maximum]   : &       1.41&(0.040)&        -0.44&(0.13)&           1.3&(0.44)&          1.59&(0.089)&        -0.81&(0.43)&           1.3&(0.48)&          0.91&(0.58)\\  
    \hline
  \end{tabular}
  \normalsize
  \caption{%
    Fit parameters for centre-to-limb variation for various lines from USUN
    and (in the case of the \ion{O}{v} 630 \AA\ line) SYNOP mosaics.  Standard
    deviations are in parentheses. The \ion{O}{iv} 554~\AA\ radiances refers
    to the sum of all the fine structure components.
    The quantities $I^\mathrm{p}(1)$ and $I^\mathrm{q}(1)$ are centre-disk
    radiances estimated from power-law and quadratic fits respectively
    (photons s$^{-1}$ cm$^{-2}$ arcsec$^{-2}$).
    The $f^\mathrm{p}_\mathrm{cl}$ and $f^\mathrm{q}_\mathrm{cl}$ are the
    correction factors to obtain  the mean disk
    radiance from the centre-disk radiances 
      (Eq.~\ref{eq:fcl}), for the quadratic (Eq.~\ref{eq:c2l_pol}) and 
      power-law (Eq.~\ref{eq:c2l_pow})
      fits, respectively.%
  }
  \label{tab:c2l_fits}
\end{sidewaystable*}

\subsubsection{Variability of quiet Sun radiances during the activity cycle}\label{sec:disc:qs:cycle}

\begin{figure}[!hbt]
  \centering
  \includegraphics[clip=true,trim=30 28 40 40,width=\linewidth]{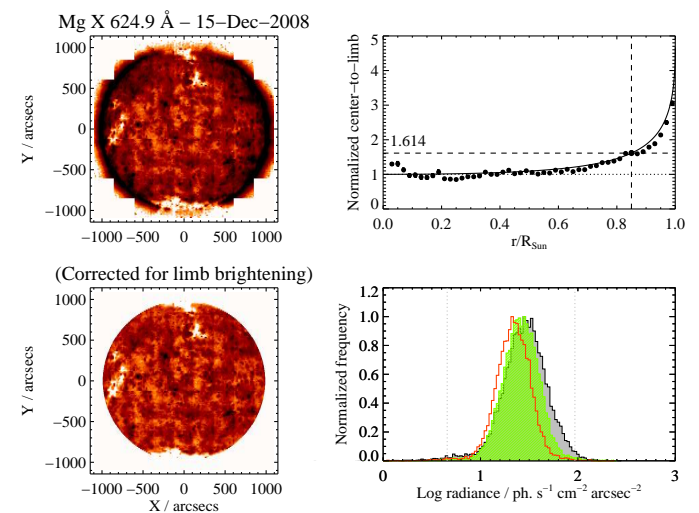}
  \caption{%
      Effect of centre-to-limb variation in the radiance histograms of the
      \ion{Mg}{x} 625~\AA\ line.  
      The original radiance map from the USUN mosaic taken on December 15,
      2008 is shown in the upper-left panel.
      The fit to the average centre-to-limb variation at the minimum of solar
      activity is shown in the upper-right panel as a solid line (i.e.:
      $a=0.23$, $b=0.44$); for comparison, the measured variation for that
      date (dots) is also shown, along with the value of limb brightening at
      $\rho=0.85$ (dashed lines).
      The map of radiances corrected for that limb brightnening function is
      shown in the lower-left panel.  
      The lower-right panel shows the radiance histograms formed from all
      on-disk pixels (grey-shaded histogram), as compared with the histogram from
      pixels with $\rho<0.85$ (green-shaded histogram); the histogram from the
      radiances corrected for the centre-to-limb variation is shown in red.
      The widths of the Gaussian fit to the core of the histograms are,
      respectively: 0.21, 0.19, and 0.17.
  }
  \label{fig:hist_c2l}
\end{figure}

Using the characterisation of the centre-to-limb variation in line median
radiances described in the previous section, we first
demonstrate in Fig.~\ref{fig:hist_c2l} how the
  radiance histograms in some lines over the full disk are biased in both the
  position of the peak and the width, if the centre-to-limb variation is
  not taken into account.  We chose to show the \ion{Mg}{x} 625~\AA, as a
  representative case of a coronal line with a clearly defined QS
  component.

  Such biases in peak position and width of the QS histograms are still
  clearly detectable when restricting the analysis to pixels within
  $0.85\times R_\mathrm{Sun}$ (corresponding to $\mu= 0.527$ or $\sim
  70$\% of the disk): this is not surprising since the value of at $\rho=0.85$
  of the centre-to-limb function is in this case $\sim 1.6$ (upper-right
  panel of Fig.~\ref{fig:hist_c2l}).  Even when restricting the analysis to
  just 50\% of the disk, or $r<0.7\times R_\mathrm{Sun}$, the bias is still
  detectable, although to a lesser extent.  In the latter case, however, the
  statistics are significantly worsened because both of a smaller number of
  pixels contributing to the histogram and a larger contribution
  from active region belts (outside the epochs of minimal solar activity).

  We chose therefore to consider all on-disk pixels but removed
  the average centre-to-limb variations.
  The resulting images were then analysed to determine the quiet-Sun component
  of the radiance histogram.

\begin{figure*}[!htbp]
  \centering
  \includegraphics[clip=true,trim=0 25 30 35,width=0.95\linewidth]{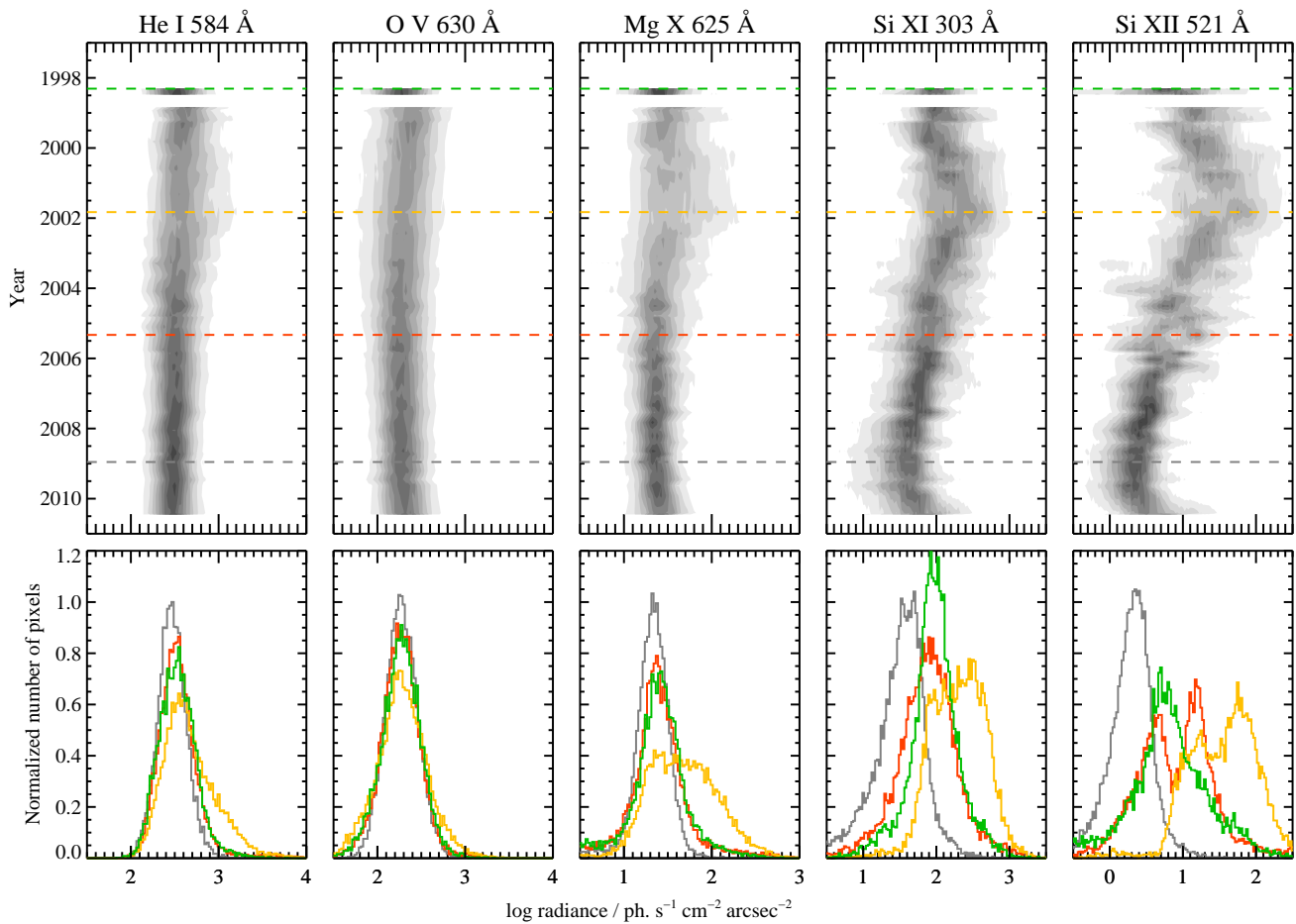}
  \caption{%
      The contours in the upper panels show the time variability of full-disk
      histograms from CDS USUN mosaics after correction from centre-to-limb
      variation.
      The horizontal lines mark representative dates: 23 April 1998 (green), 30
      October 2001 (orange), 1 May 2005 (red), and 15 December 2008 (grey). The gap in
      1998 is due to the loss of contact of SOHO.
      The lower panels show the corresponding radiance histograms with the
      same colour coding.   
      All the histograms are normalised to the total number of pixels; those
      shown in the lower panels are then also normalised 
      to the peak of the average histogram at the solar minimum of activity.
    }
  \label{fig:hst_timeline}
\end{figure*}

In the upper panels of Fig.~\ref{fig:hst_timeline}, we show 
the evolution of radiance histograms along the cycle, once the centre-to-limb
variation is removed.  
It is readily apparent that 
the quiet Sun contribution is usually still well
approximated by a log-normal distribution, even around the maximum of the
cycle.  
We therefore estimated the position of the peak and its width by fitting a Gaussian
to histogram bins around the maximum of the distribution.  In case of
multi-component distributions, we fitted the peak corresponding to the lowest
mean radiance.

In the lower panels of the same figure, we show examples of radiance
histograms at different phases of the solar cycle for four representative
dates from the rising phase of cycle 23 (23 April 1998: green) through its
maximum (30 October 2001: orange) and from the decaying phase (1 May 2005: red), to the
deep minimum before cycle 24 (15 December 2008: grey histogram).
With the exception of the \ion{Si}{xii} 521~\AA\ and
of the \ion{Si}{xi} 303~\AA\ histograms, the fitted component in these
examples corresponds very well to the
component observed during the 2008 minimum and, therefore, can be interpreted
as a QS component.

In the \ion{Si}{xii} 521~\AA\ histograms, on the other hand, 
even the lowest-radiance components do not
  overlap with the quiet component measured at solar minimum.
  For example, the lowest component in the 2005 (red) histogram appears at
  around $\log I \sim 0.6$ -- $0.7$ (radiance, $I$, in photons s$^{-1}$
  cm$^{-2}$ arcsec$^{-2}$) at about $0.3$--$0.4$ dex higher than the 2008 peak.
  A QS component at that position is seen only as a ``hump'' above the
  Gaussian fit.  A similarly small contribution is also seen in the April 2008
  histogram too but is almost completely absent in the histogram
  corresponding to the maximum of solar activity (October 2001).
  Similar considerations can be made for the \ion{Si}{xi} 303~\AA\ histograms,
  although the overlap with the histogram at the minimum of activity remains
  significant even at the peak of the cycle.

  This analysis highlights a feature of the solar EUV emission during its
  activity cycle: while the radiance in lines that form at temperatures lower
  than $\log T \sim 6$ still shows a noticeable emission with mean
  radiances which are comparable to QS values during the solar minimum of activity, 
  such a component is strongly decreased in hotter lines to the point of
  virtually disappearing in the hottest lines in the CDS/NIS range ($\log T \gtrsim 6.3$) at epochs near the solar maximum.

\begin{figure*}[!htbp]
  \centering
  \includegraphics[clip=true,trim=0 40 40 40,width=0.45\linewidth]{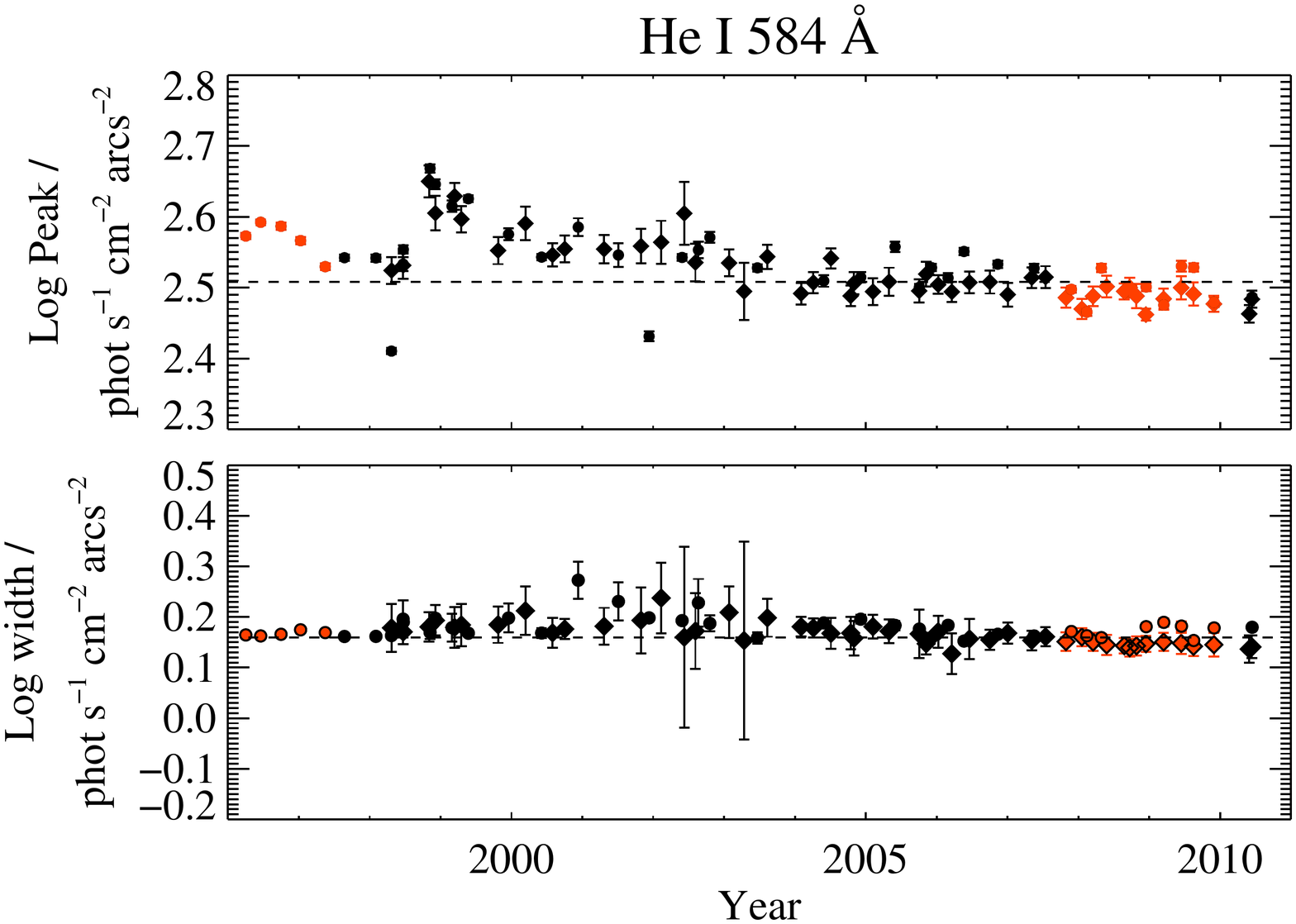}
  \includegraphics[clip=true,trim=0 40 40 40,width=0.45\linewidth]{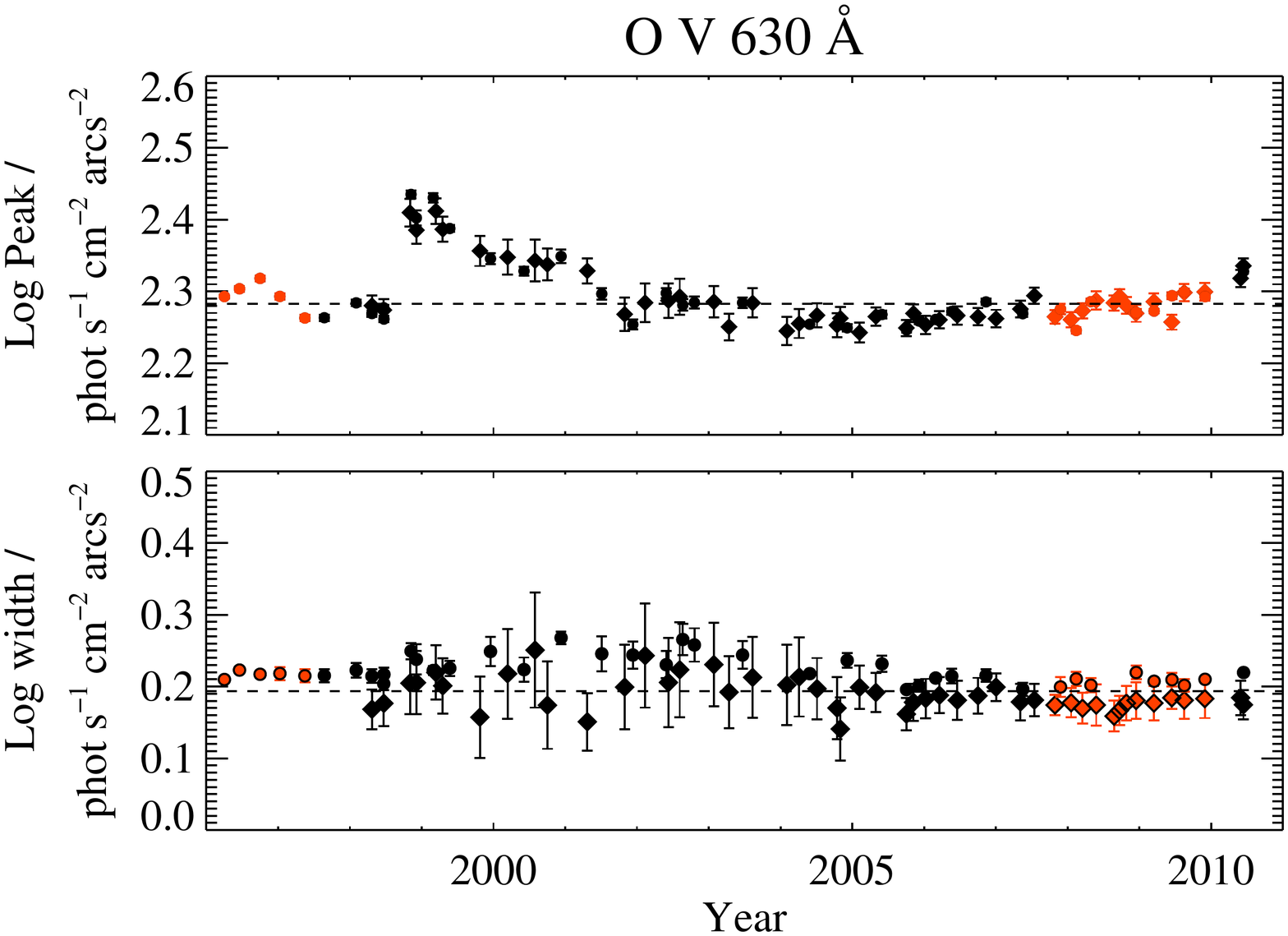}\\
  \includegraphics[clip=true,trim=0 40 40 40,width=0.45\linewidth]{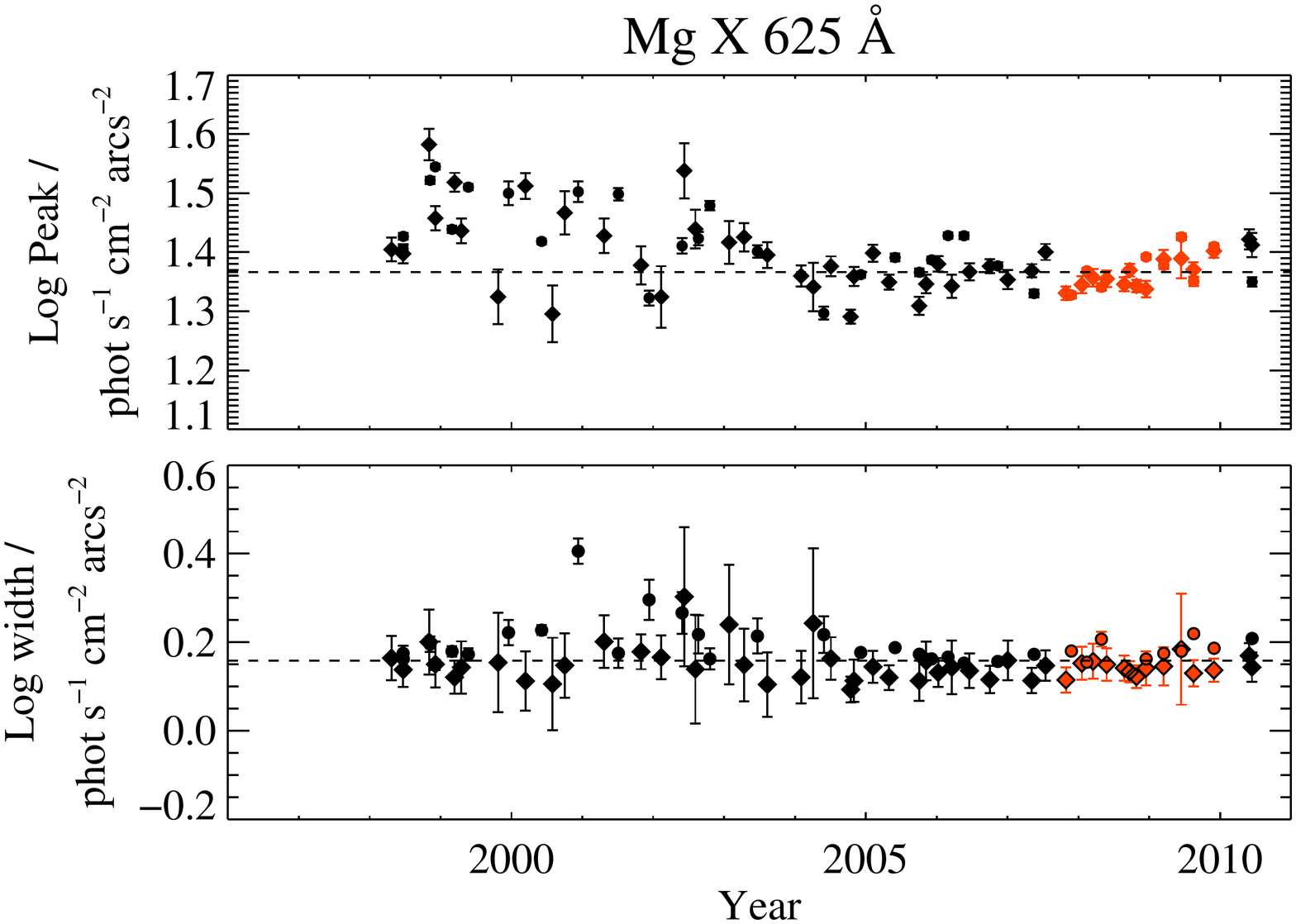}
  \includegraphics[clip=true,trim=40 40 0 40,width=0.45\linewidth,angle=180,origin=c]{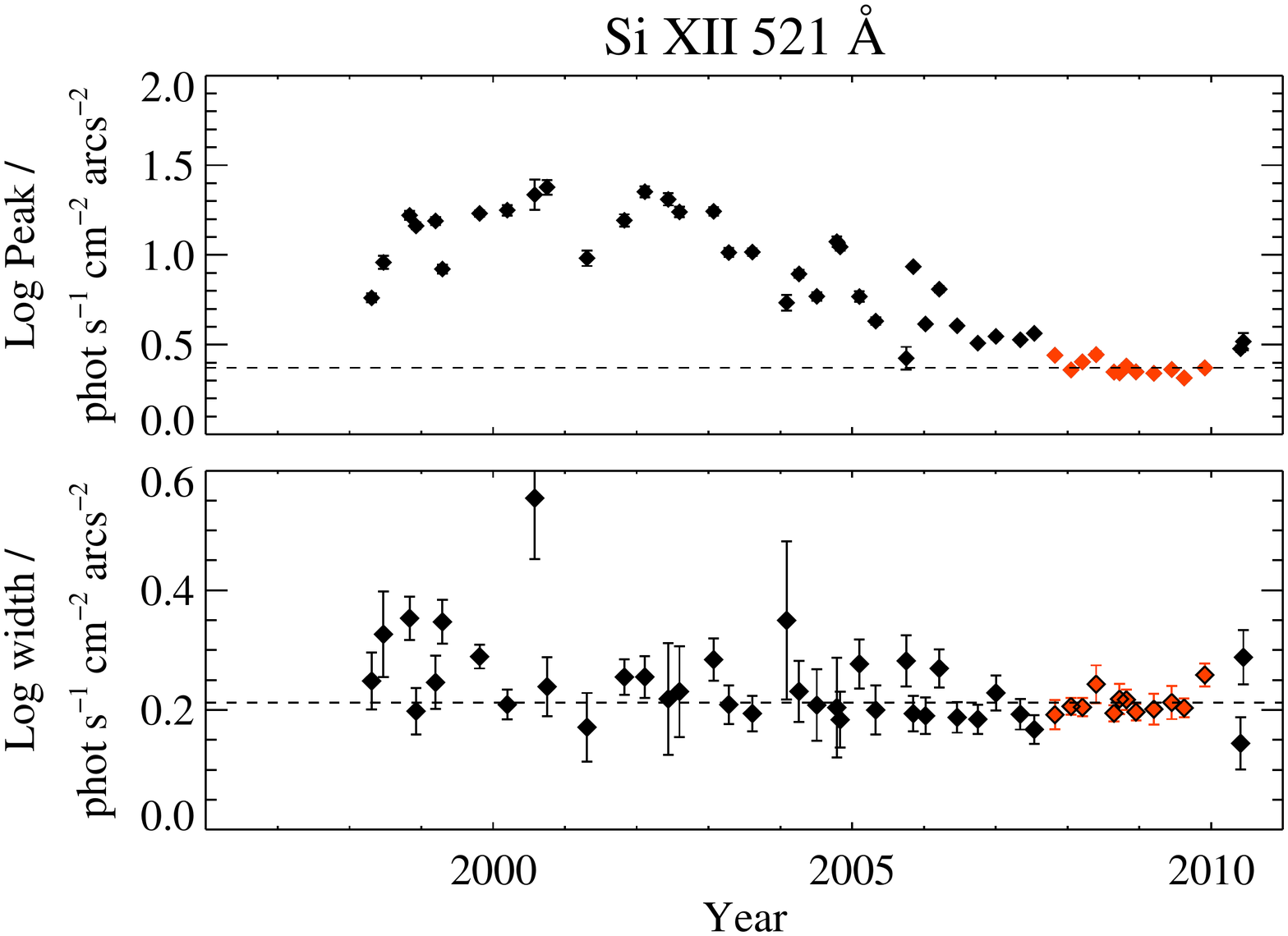}\\
  \caption{%
    Fit parameters of the QS or lowest-radiance component of the radiance
    histograms for four representative lines.  Filled diamonds and circles denote USUN
    and SYNOP mosaics, respectively.  The data points used to compute the
    average values of Table~\ref{tab:qs_params} are marked in red.
  }
  \label{fig:hist_fits}
\end{figure*}

A more detailed characterisation of the variability in the distribution of QS
radiances along the solar cycle  
is given in Fig.~\ref{fig:hist_fits},
which 
shows how the position of the peak and width of the quiet Sun or of the 
lowest-radiance component
of the radiance histogram vary with the cycle for four lines.  
The measurements shown in the figure for the \ion{He}{i} 584~\AA, \ion{O}{v}
630~\AA, and \ion{Mg}{x} 625~\AA\ lines are from both USUN (diamonds) and
SYNOP (circles) mosaics.
However, only the \ion{He}{i} 584~\AA\ and \ion{O}{v} 630~\AA\ lines (as well
as the \ion{Mg}{ix} 368~\AA\ and \ion{Fe}{xvi} 361~\AA\ lines, not shown here)
were included in SYNOP studies, since the beginning of the SOHO mission and
thus cover both the 1996-1997 and the 2007-2009 minima; the \ion{Mg}{x}
625~\AA\ line was included in SYNOP mosaics only after mid-1998 during
the rising phase of cycle 23.

  It is readily apparent how the peak of the radiance distribution of cooler lines
  is nearly constant throughout the solar cycle, except for the period
  1998--2002 (or 1998--2004, for the \ion{Mg}{x} 625~\AA\ line), where the
  distributions of TR lines are perturbed by the signal due to active
  regions. This is not easily disentangled from the QS component (see again
  histograms of Fig.~\ref{fig:hst_timeline}).

  The lowest-radiance component of the \ion{Si}{xii} 521~\AA\ distribution, on
  the other hand, is systematically higher by a factor 3 to an order of
  magnitude relative to the well-defined value at solar minimum.  As shown in
  at least two of the examples of Fig.~\ref{fig:hst_timeline}, a true QS
  component may still be seen in the radiance histograms, except at the peak
  of solar activity, but this component is overcome by other components
  associated with ARs.  Hence, the variability shown in Fig.~\ref{fig:hist_c2l}
  of the lowest-radiance peak is to be attributed to the appearance of ARs
  along the cycle, not to an intrinsic variability of the QS component.  In
  principle, with careful fitting of multiple Gaussians, or masking out
  the ARs on the disk, it could be possible to recover the properties of the
  residual QS component.  Such an approach, however, is made difficult by the
  large area covered by AR-related emission, as discussed in
  Sec.~\ref{sec:disc:irr}.  For the purpose of this paper, the current
  analysis is sufficient.

  A remarkable feature common to all these radiance components, on the
  other hand, is their
  constant width, even in the case of the \ion{Si}{xii} 521~\AA\ line.
  This is true for the QS or lowest-radiance component but
  is also true for the components due to the emission by active regions (see again
  Fig.~\ref{fig:hst_timeline}) to some extent.  It is beyond the scope of this paper to
  analyse the properties of the histogram components due to ARs in detail, but
  it seems the value of these histogram widths are
  characteristic of each line from the data shown here.  

  \begin{table}[!htb]
    \centering
    \begin{tabular}[tb]{lr@{\,}lr@{\,}l}
      \hline
      \hline
      Line & 
      \multicolumn{2}{c}{Peak}    & 
      \multicolumn{2}{c}{Width} \\
      \ion{He}{i} 584 \AA     &       2.51& (0.037)  &   0.159& (0.014) \\
      \ion{He}{i} 537 \AA     &       1.55& (0.013)  &   0.144&(0.0069) \\
      \ion{He}{i} 522 \AA     &      0.981& (0.013)  &   0.146&(0.0080) \\
      \ion{He}{i} 516 \AA     &      0.751& (0.031)  &   0.144& (0.011) \\
      \ion{He}{ii} 304 \AA    &       3.22& (0.016)  &   0.143&(0.0079) \\
      \ion{O}{iii} 600 \AA    &       1.39&(0.0089)  &   0.187& (0.011) \\
      \ion{O}{iv} 554 \AA     &       2.07& (0.017)  &   0.178&(0.0090) \\
      \ion{O}{v} 630 \AA      &       2.28& (0.017)  &   0.194& (0.020) \\
      \ion{Mg}{x} 625 \AA     &       1.37& (0.028)  &   0.159& (0.028) \\
      \ion{Si}{xi} 303 \AA    &       1.62& (0.039)  &   0.239& (0.023) \\
      \ion{Si}{xii} 521 \AA   &      0.371& (0.040)  &   0.212& (0.020) \\
      \hline
    \end{tabular}
    \caption{%
      Average fit parameters for QS component at solar minimum for the same
      lines as Table~\ref{tab:c2l_fits}
      using both  USUN and SYNOP mosaics.  Logarithm of radiances in
      units of photons s$^{-1}$ cm$^{-2}$ arcsec$^{-2}$.
    }
    \label{tab:qs_params}
  \end{table}

Table~\ref{tab:qs_params} reports the average values at solar-minimum epochs,
as defined in Sec.~\ref{sec:disc:qs:c2l}, of both the position of the
peak and the width of the QS component in the radiance distribution.  The
data points used to compute the averages are highlighted in red in
Fig.~\ref{fig:hist_c2l}.  We recall that the position of the peak is in
practice a measure of the \emph{median} radiance; the \emph{mean} radiance can
be computed by using Eq.~\ref{eq:bias}.  These average values were computed
by averaging values from both USUN and, whenever available, SYNOP mosaics: from
the discussion of Sec.~\ref{sec:disc:hist:comp}, the different sampling size
and resolutions of the two types of mosaics do not affect the basic properties
of the radiance histograms.  This is also clearly seen from the data in
Fig.~\ref{fig:hist_fits}.

As already mentioned in Sec.~\ref{sec:disc}, earlier analysis of radiance
histograms in terms of log-normal distributions were done by
\citet{Wilhelm-etal:98} and \citet{Fontenla-etal:07}: they also
report measurements for the positions and widths of the histograms.

There is no overlap in the set of lines analysed from SUMER spectra by
\citet{Wilhelm-etal:98}, as noted in Sec.~\ref{sec:disc:qs:c2l}, 
with the exception of the $2^\mathrm{nd}$ order lines
\ion{He}{i} 584~\AA\ and \ion{O}{v} 630~\AA.  They found the width of the
radiance histogram of the \ion{He}{i} line to be about 0.2 dex (their Table~3),
while we measure a value of $0.16 \pm 0.015$ dex (Table~\ref{tab:qs_params}); in
the case of the \ion{O}{v} line, they found that its histogram has a width of about
0.26 dex, while we measure $0.19 \pm 0.02$ dex.  They also measured the width of
the distribution of the \ion{O}{iv} 788~\AA: their values range from 0.275 to
0.30 dex, while we find $0.178 \pm 0.01$ dex for the
\ion{O}{iv} 554~\AA\ multiplet of the same ion.

It is not clear why 
the values measured by \citet{Wilhelm-etal:98} are generally higher than those
found here.  We  note
that the procedure adopted by those authors apparently involves a Gaussian fit
of the full radiance histograms, which, as we repeatedly noted in this work,
may include ``humps'' in the wings due to features different from the QS (CHs,
DHs, ARs; see also their Fig.~8).  We were very careful in fitting
only the core of the histograms, thus filtering out most other contributions.
Moreover, the values measured by \citet{Wilhelm-etal:98} were derived from
only a very few disk scans (one to four for each line) and, indeed, usually
show a considerable scatter in the widths of their histograms (see their
Fig.~9), while our tabulated values refer to an average over a couple of
dozens of measurements.

For completeness we report here the values obtained by
\citet{Fontenla-etal:07}: they find the widths of the distributions of the
1430~\AA\ continuum, the Ly-$\alpha$ line, and the \ion{Ca}{ii} K$_3$
filtergrams to be 0.245, 0.271, and 0.202 dex, respectively.  Despite the fact
that those features are formed in different regions of the atmosphere
than the CDS EUV lines, the similar widths of their radiance distributions is
noteworthy and, perhaps, points to a close connection between the energy
dissipation mechanisms in those atmospheric regions.

From the results presented here, it is clear that this distribution for either the
cooler TR lines or coronal lines that form at $\log T < 6.2$, is not
significantly altered along a solar cycle.  Even for lines forming at $\log T
> 6.2$, it is more likely that a QS component similar to the one measured at
epochs of minimum of solar activity is still present but is overcome by the
appearance of a much stronger and diffuse emission due to ARs.

This result contradicts the findings of \citet{Kamio-Mariska:12}: from the
analysis of Hinode spectra, they find the QS emission to be
intrinsically variable during the time interval they considered (end of 2006
to early 2011).  

Even our data, such as those shown in Fig.~\ref{fig:hist_c2l}, could be still
be compatible with a variability of the QS median radiance (but not of the
width of the distribution) during the maximum of solar activity.  However,
such a variability would be rather small -- of the order at most of 0.1--0.2 dex for
the \ion{Mg}{x} 625~\AA\ line; the variation found by \citet{Kamio-Mariska:12}
is of the order of a factor between 2 and 5 for lines of similar formation temperature
(like the \ion{Fe}{xi} 180~\AA\ line) or hotter lines.  In any case, our data for \ion{Mg}{x}
clearly show that the median radiance settles to a constant value by the end
of 2004, which is much earlier that the Hinode measurements.

One obvious explanation of the discrepancy is  that 
\citet{Kamio-Mariska:12} used a set of single-slit exposures near the 
disk centre. That is, their measurements are affected by a very small field
of view and the occasional presence of active regions. 
Another problem in the \citet{Kamio-Mariska:12} results is related to their
radiometric calibration, which is now known to be at fault.
\cite{DelZanna:13} produced a completely new Hinode EIS calibration
and applied it to a careful selection of the same dataset  used by 
\citet{Kamio-Mariska:12} in their analysis. The \cite{DelZanna:13}  
results are that the quiet Sun radiances in lines formed at or below 1~MK 
do not change, which agree with the present results.

We also note that the conclusions of
\citet{Kamio-Mariska:12} rely heavily on the comparison of radiance histograms
taken near the disk centre at two specific epochs: December 2006 and February
2009.  In particular, they find that the peak of the radiance histogram of the
\ion{Fe}{xii} 194~\AA\ line is at $\sim 200$ ergs cm$^{-2}$ s$^{-1}$
sr$^{-1}$ in 2006 and $\sim 70$ ergs cm$^{-2}$ s$^{-1}$ sr$^{-1}$ in 2009
(see their Fig.~3).

We do not have USUN scans for December 2006, but we do show in
Fig.~\ref{fig:img_timeline} (discussed in Sec.~\ref{sec:disc:irr}) scans for 2
October 2006, and 2 January 2007, which both show some significant activity at the
equator with a broad longitudinal distribution.  Indeed, inspection of EIT
images around mid-December 2006 (e.g.: 12 December 2006) reveals that one active
region near the equator (NOAA 10930) exhibits
some long-range loops extending to a plage-like, un-numbered trailing area, at least in EIT~195 images.

We therefore speculate that the mean radiance estimated by
\citet{Kamio-Mariska:12} for December 2009 may be contaminated by solar
activity.  We notice that the
radiance histogram for that epoch has a ``hump'' corresponding to the main
component of the more quiescent 2009 histogram in their Fig.~3.  That is perfectly analogous
to what we find, for instance, in some of the histograms of our
Fig.~\ref{fig:hst_timeline}, as discussed earlier in this section.

\subsection{On the variability of the EUV solar irradiance}\label{sec:disc:irr}

\begin{figure*}[!htbp]
  \centering
  \includegraphics[clip=true,trim=-10 110 70 130,width=\linewidth]{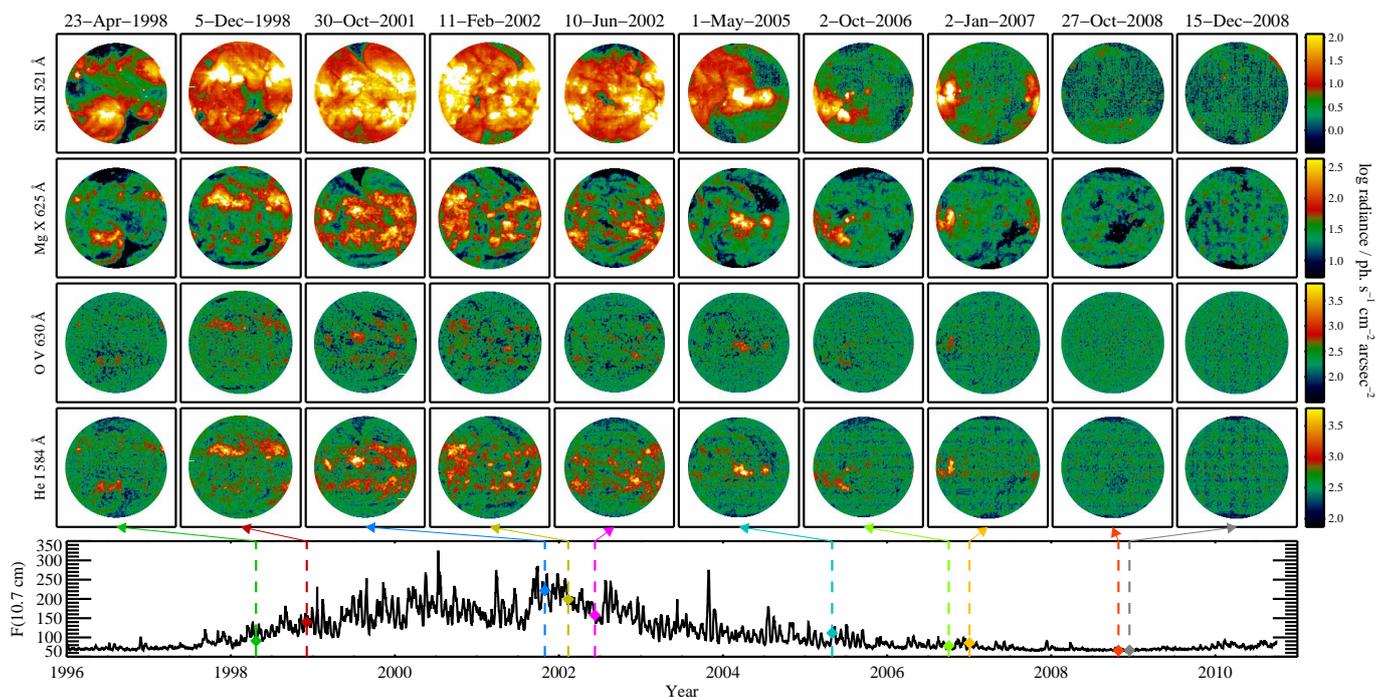}
  \caption{%
    Maps of line radiance from USUN mosaics in four lines (top to bottom rows): \ion{Si}{xii}
    521~\AA, \ion{Mg}{x} 625~\AA, \ion{O}{v} 630~\AA, and \ion{He}{i}
    584~\AA.  The dates of the USUN mosaics are marked by vertical, dashed
    lines in the lowest panel, showing the variability of the radio flux at
    10.7 cm.  
  }
  \label{fig:img_timeline}
\end{figure*}

A remarkable result described in \citetalias{DelZanna-Andretta:11} is that solar
EUV spectral irradiance due to TR lines varies by only modest amounts during
the solar cycle (less than a factor 2 for the \ion{O}{v} 630~\AA\ line). This is in
contrast with the strong variability of coronal lines (at least one order of
magnitude for the \ion{Si}{xii} 521~\AA\ line).  Even excluding the off-limb
emission, as estimated to be non-negligible for hotter lines in
\citetalias{DelZanna-Andretta:11}, this different behaviour calls for an
explanation.  The analysis of the previous section of the properties of the
statistical distribution of radiances may turn out useful for studying that phenomenon more in
detail.

Figure~\ref{fig:hst_timeline} already provides some clues to the present
discussion.  We already noted how the contribution of AR emission relative to
the QS is significantly larger in hotter lines.  At the peak of solar
activity, the QS contribution virtually disappears from the distribution of
hottest lines (like the \ion{Si}{xii} 521~\AA\ line), in contrast to the TR
lines whose AR contribution always remains a mere perturbation to the main QS
distribution.  We also note that the AR contribution in hotter lines can
easily peak at radiances at least one order of magnitude higher than the QS
median value at solar minimum; conversely, the bulk of AR emission in TR lines
always remains within a factor 2--3 from the the QS peak of the distribution.

These observations are illustrated in Fig.~\ref{fig:img_timeline}, which shows a set
of radiance maps in some CDS lines at representative times of the solar cycle.
The figure is analogous to Fig.~1 of \citetalias{DelZanna-Andretta:11}, except
that the centre-to-limb variation for each line has been removed and that the
colour table has been chosen to distinguish quiescent regions from
emission related to solar activity.  The set of images shown in this figure
includes those from the same dates as shown in Fig.~\ref{fig:hst_timeline}
plus six others for a more complete coverage of the cycle.  
Along with the radiance maps, we also show the variability of solar flux at
2800~MHz (10.7~cm) for which we have adopted the daily measurements (scaled
to 1 AU) made by the National Research Council of Canada (at the Algonquin
Radio Observatory near Ottawa until 1991 and at the Dominion Radio
Astrophysical Observatory near Penticton, British Columbia).

Even a casual inspection shows that regions appearing in a darker colour in
\ion{Mg}{x} 625~\AA\ images (or in cooler lines) are bright in the hotter
\ion{Si}{xii} 521~\AA\ line and are definitely brighter than typical QS
emission in that line during solar minimum.  In
particular, the red-yellow areas correspond to line radiance above the peak,
$p$, of QS radiance given in Table~\ref{tab:qs_params} by at least
$2w$, where $w$ is the width of the log-normal QS distribution
(third column of Table~\ref{tab:qs_params}).  Thus, there is a probability of
2.3\% to find values belonging to the QS log-normal distribution above that
threshold.

At the peak of the solar cycle (e.g.\ in 2001), nearly 100\% of the solar disk
emits in \ion{Si}{xii} 521~\AA\ above that threshold.  Only CHs tend to remain
as dark as at the minimum of activity: the QS emission nearly disappears.  By
contrast, spectra with \ion{O}{v} 630~\AA\ radiance above the threshold are
less than 12\% of the total over the solar disk, even at the peak of solar
activity.  Considering that 2.3\% of the spectra belong to the tail of the QS
log-normal distribution, only the remaining 10\% or less is due to AR
emission.

The \ion{He}{i} and \ion{He}{ii} lines deviate from the numbers typical of TR lines:
the number of spectra with radiances above the threshold can reach about 25\%
of the disk.  This can be explained by the effect of enhancement of
those lines due coronal back-radiation from the nearby ARs or by some other
enhancement mechanism differentially operating on the lines of that element
\citep[e.g.:][]{Andretta-etal:00,Andretta-etal:03}.

\begin{figure}[!htbp]
  \centering
  \includegraphics[clip=true,trim=20 45 55 55,width=\linewidth]{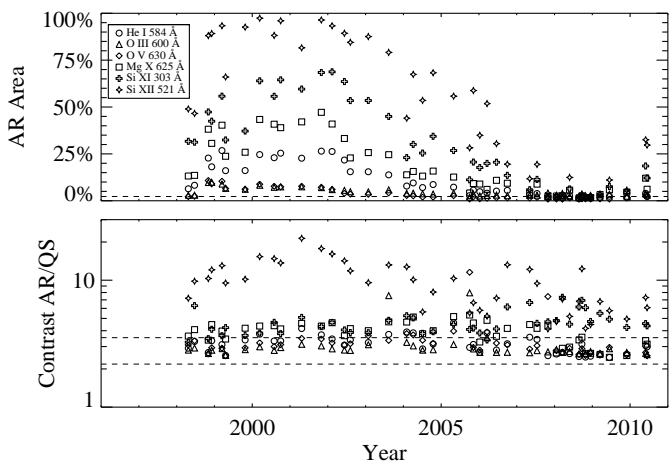}
  \caption{%
    Estimated AR area (top panel) and contrast relative to the QS (bottom
    panel) during solar cycle 23 for various lines: %
    \ion{He}{i} 584~\AA\ (circles), %
    \ion{O}{iii} 600~\AA\ (triangles), %
    \ion{O}{v} 630~\AA\ (diamonds), %
    \ion{Mg}{x} 625~\AA\ (squares), %
    \ion{Si}{xi} 303~\AA\ (crosses), and %
    \ion{Si}{xii} 521~\AA\ (stars).
  }
  \label{fig:AR_area}
\end{figure}

In Fig.~\ref{fig:AR_area}, we show how the statistical properties of the
radiance histograms above the $p+2w$ threshold vary with the solar
cycle, as compared to the bulk of the QS log-normal distribution defined by the
parameters of Table~\ref{tab:qs_params}.

The upper panel shows the fraction of disk spectra with radiance in a given
line that is higher than the threshold.  That number, minus the 2.3\% contribution due
to the tail of the QS distribution (dashed, horizontal line), can be
interpreted as an estimate of the fraction of the solar disk covered by ARs at
a given epoch and seen in a given line.

The lower panel shows the ratio between the mean value above the threshold
$p+2w$ (mostly AR contribution) and the mean within the interval $p
\pm 2w$ (which includes 95.5\% of the QS contribution and should exclude
most of CH and DH contributions).  We refer to this ratio as to ``AR/QS
contrast''.  For log-normal distributions of width 0.15 and 0.25 dex, that
ratio is 2.2 and 3.5, respectively; 
those values, shown as horizontal, dashed lines, 
can be considered as lower limits for the actual measurements.

From the upper panel, the ``AR area'' as function of line formation
temperature appears to be roughly constant for TR
lines (with the exception  of \ion{He}{i} and \ion{He}{ii} lines, see below).  For
coronal lines, the AR areas increase rapidly instead and yet smoothly, where it is up to
almost 100\% for lines forming at $T>2$~MK, as noted above.  This is not
entirely surprising: while TR emission is expected to come from a
narrow, albeit inhomogeneous layer near the solar surface, emission of hotter
lines is expected to come mostly from larger AR or inter-AR loops.  Thus, we
speculate that this ``AR'' component is due either to a diffuse corona due
to an enhanced activity or to a large number of overarching loops joining different
active regions (the former can be due to the latter).

More intriguing is what is shown in the lower panel: the contrast betwen AR
and QS remains between a factor 2 and 5 for all the lines up to the
\ion{Si}{xi} 303~\AA\ (forming at $T\sim 1.5$~MK), while suddenly jumping
to an order of magnitude or more (up to a factor 30) for the \ion{Si}{xii}
521~\AA\ line (forming at $T\sim 2$~MK) or for hotter lines, such as the
\ion{Fe}{xvi} 361~\AA, which is not shown in the plots.
The same effect can be seen in the histograms of Fig.~\ref{fig:hst_timeline}:
the positions of the main peaks of the \ion{Si}{xi} 303~\AA\ radiance
distributions remain within a factor 10 or less during the cycle, while
it varies by at least a factor 30 between minimum and maximum for the
\ion{Si}{xii} 521~\AA\ line.

Such an abrupt change of AR mean emission compared to QS between the formation
temperatures of \ion{Si}{xi} and \ion{Si}{xii} may signal the appearance at
solar maximum of a hot component of the corona missing during the solar
minimum.  This tentative interpretation needs, however, to be addressed more
carefully, including a quantitative comparison with X-ray measurements, for
instance, which is a task beyond the scope of this work.

\section{Conclusions}\label{sec:end}

The database of CDS/NIS synoptic observations (SYNOP and USUN) taken during
the lifetime of the instrument has been proven in
\citetalias{DelZanna-Andretta:11} to be very useful in the study of the solar EUV
irradiance during solar cycle 23.  In this work, we continued the analysis of
the data by studying the properties of the solar EUV radiances of a number of
strong lines. This is the first time that such an analysis is possible; 
hence, all the present results have particular importance.

The main points we addressed in this paper can be summarised as follows:
\begin{itemize}
\item We showed that structures around active regions, sometimes referred to as
  dark halos or dark canopies are common, and discussed their
  similarities and differences with coronal holes.  In particular, we show how
  they are well visible in TR lines, contrary to coronal holes.
\item We find that the limb-brightening for all the EUV lines in the CDS/NIS
  range is consistent with small or negligible optical thickness with the
  notable exception of the \ion{He}{i} and \ion{He}{ii} lines.
\item The limb-brightening significantly affects any characterisation of 
the solar radiances. 
\item The limb-brightening function for all the lines does not change
  measurably during the cycle. In particular, the limb-brighenining was not
  different in 2008 than in 1996 in the only line, \ion{O}{v} 630~\AA, for
  which a comparison was possible. Our limb-brightening curves differ 
from those previously published.
\item We confirm earlier findings that the radiance histogram have a well
  defined, log-normal QS component, which is almost always identifiable in full-disk
  distributions, especially in coronal lines.
However, our results differ from previous ones.
\item The width of the lowest-radiance, log-normal distribution (either QS or
  from ``diffuse'' corona) is constant along the cycle.
\item At solar maximum, the QS component is strongly reduced or even vanishes
  in lines with formation temperature $T > 2$~MK.  More accurately, the QS
  component is progressively overcome by a much larger contribution from ARs.
\item We analysed the variability of solar irradiance in terms of variable
  area and ``intrinsic'' emission of active regions.  We found that the
  emission from active regions in TR and coronal lines forming at $T<1.5$~MK
  undergo only relatively modest changes during the cycle: thus, much of the
  irradiance variability is due to a change in the emitting area.  For hotter
  lines (i.e.\ lines forming at $T>2$~MK), the emitting area saturates to
  almost 100\% of full solar disk at the maximum of activity, while
  the emission due to active regions simultaneously increases by more than an
  order of magnitude.
\end{itemize} 

The analysis of both the centre-to-limb variation and of the radiance
statistical distribution points at a constant QS emission along solar cycle
23. This result is contrary to earlier claims \citep[e.g.:][]{Kamio-Mariska:12}.  Since 
both the centre-to-limb function and the widths of the radiance distributions
do not depend on the radiometric calibration of the data, we
are confident that our results point to truly constant properties of the quiet
Sun during the activity cycle.

Finally, we remark that the availability of a good radiometric calibration,
including its long-term variability, as estimated in
\citetalias{DelZanna-etal:10}, has been the key for the analysis of the
variability of EUV emission in various lines across the solar disk and during
the cycle.  
For instance, there were histogram components of some lines, which would have
been classified as ``quiescent'' on the basis of a morphological analysis only
(at the low resolution of this data set), even during solar maximum.
On the other hand, the calibrated radiances of those components turned out to be much brighter than QS
areas during solar miminum and, thus, had to be considered as major
contributors to the increased irradiance in that line during solar maximum.

\begin{acknowledgements}
GDZ  acknowledges support by   STFC (UK). 
VA acknowledges partial support by the Italian Space Agency (ASI), through
ASI-INAF contracts I/035/05/0 and I/05/07/0.
CDS was built and operated by a consortium led by the 
Rutherford Appleton Laboratory (RAL), which includes
UCL/Mullard Space Science Laboratory, NASA/ Goddard 
Space Flight Center, Max Planck Institute for Extraterrestrial 
Physics, Garching, and Oslo University.
SOHO is a mission of international cooperation between 
ESA and NASA.
The results obtained here could not have been 
achieved without the efforts of the
operation and scientific staff, which successfully 
ran  the CDS instrument, in large 
part  at STFC/RAL (UK) and NASA/GSFC (US).
\end{acknowledgements}


\begin{thebibliography}{35}
\expandafter\ifx\csname natexlab\endcsname\relax\def\natexlab#1{#1}\fi

\bibitem[{{Andretta} {et~al.}(2003){Andretta}, {Del Zanna}, \&
  {Jordan}}]{Andretta-etal:03}
{Andretta}, V., {Del Zanna}, G., \& {Jordan}, S.~D. 2003, \aap, 400, 737

\bibitem[{{Andretta} {et~al.}(2000){Andretta}, {Jordan}, {Brosius}, {Davila},
  {Thomas}, {Behring}, {Thompson}, \& {Garcia}}]{Andretta-etal:00}
{Andretta}, V., {Jordan}, S.~D., {Brosius}, J.~W., {et~al.} 2000, \apj, 535,
  438

\bibitem[{{Curdt} {et~al.}(2008){Curdt}, {Tian}, {Teriaca}, {Sch{\"u}hle}, \&
  {Lemaire}}]{Curdt-etal:08}
{Curdt}, W., {Tian}, H., {Teriaca}, L., {Sch{\"u}hle}, U., \& {Lemaire}, P.
  2008, \aap, 492, L9

\bibitem[{{Del Zanna}(2013)}]{DelZanna:13}
{Del Zanna}, G. 2013, \aap, 555, A47

\bibitem[{{Del Zanna} \& {Andretta}(2006)}]{DelZanna-Andretta:06}
{Del Zanna}, G. \& {Andretta}, V. 2006, in ESA Special Publication, Vol. 617,
  SOHO-17. 10 Years of SOHO and Beyond

\bibitem[{{Del Zanna} \& {Andretta}(2011)}]{DelZanna-Andretta:11}
{Del Zanna}, G. \& {Andretta}, V. 2011, \aap, 528, A139

\bibitem[{{Del Zanna} {et~al.}(2005){Del Zanna}, {Andretta}, \&
  {Beaussier}}]{DelZanna-etal:05}
{Del Zanna}, G., {Andretta}, V., \& {Beaussier}, A. 2005, \memsai, 76, 953

\bibitem[{{Del Zanna} {et~al.}(2010){Del Zanna}, {Andretta}, {Chamberlin},
  {Woods}, \& {Thompson}}]{DelZanna-etal:10}
{Del Zanna}, G., {Andretta}, V., {Chamberlin}, P.~C., {Woods}, T.~N., \&
  {Thompson}, W.~T. 2010, \aap, 518, A49

\bibitem[{{Del Zanna} {et~al.}(2011){Del Zanna}, {Aulanier}, {Klein}, \&
  {T{\"o}r{\"o}k}}]{DelZanna-etal:11}
{Del Zanna}, G., {Aulanier}, G., {Klein}, K.-L., \& {T{\"o}r{\"o}k}, T. 2011,
  \aap, 526, A137

\bibitem[{{Del Zanna} {et~al.}(2001){Del Zanna}, {Bromage}, {Landi}, \&
  {Landini}}]{DelZanna-etal:01_cdscal}
{Del Zanna}, G., {Bromage}, B.~J.~I., {Landi}, E., \& {Landini}, M. 2001, \aap,
  379, 708

\bibitem[{{Doschek} {et~al.}(1976){Doschek}, {Feldman}, {VanHoosier}, \&
  {Bartoe}}]{Doschek-etal:76}
{Doschek}, G.~A., {Feldman}, U., {VanHoosier}, M.~E., \& {Bartoe}, J.-D.~F.
  1976, \apjs, 31, 417

\bibitem[{{Feldman} {et~al.}(2000){Feldman}, {Dammasch}, \&
  {Wilhelm}}]{Feldman-etal:00}
{Feldman}, U., {Dammasch}, I.~E., \& {Wilhelm}, K. 2000, \ssr, 93, 411

\bibitem[{{Feldman} {et~al.}(2003){Feldman}, {Dammasch}, {Wilhelm}, {Lemaire},
  {Hassler}, \& {Battrick}}]{SUMER-Atlas:03}
{Feldman}, U., {Dammasch}, I.~E., {Wilhelm}, K., {et~al.}, eds. 2003, ESA
  Special Publication, Vol. 1274, {Images of the solar upper atmosphere from
  SUMER on SOHO}

\bibitem[{{Feldman} {et~al.}(1976){Feldman}, {Doschek}, {Vanhoosier}, \&
  {Purcell}}]{Feldman-etal:76}
{Feldman}, U., {Doschek}, G.~A., {Vanhoosier}, M.~E., \& {Purcell}, J.~D. 1976,
  \apjs, 31, 445

\bibitem[{{Fontenla} {et~al.}(2007){Fontenla}, {Curdt}, {Avrett}, \&
  {Harder}}]{Fontenla-etal:07}
{Fontenla}, J.~M., {Curdt}, W., {Avrett}, E.~H., \& {Harder}, J. 2007, \aap,
  468, 695

\bibitem[{{Gallagher} {et~al.}(1998){Gallagher}, {Phillips}, {Harra-Murnion},
  \& {Keenan}}]{Gallagher-etal:98}
{Gallagher}, P.~T., {Phillips}, K.~J.~H., {Harra-Murnion}, L.~K., \& {Keenan},
  F.~P. 1998, \aap, 335, 733

\bibitem[{{Griffiths} {et~al.}(1999){Griffiths}, {Fisher}, {Woods}, \&
  {Siegmund}}]{Griffiths-etal:99}
{Griffiths}, N.~W., {Fisher}, G.~H., {Woods}, D.~T., \& {Siegmund}, O.~H.~W.
  1999, \apj, 512, 992

\bibitem[{{Harrison} {et~al.}(1995){Harrison}, {Sawyer}, {Carter}, {Cruise},
  {Cutler}, {Fludra}, {Hayes}, {Kent}, {Lang}, {Parker}, {Payne}, {Pike},
  {Peskett}, {Richards}, {Gulhane}, {Norman}, {Breeveld}, {Breeveld}, {Al
  Janabi}, {McCalden}, {Parkinson}, {Self}, {Thomas}, {Poland}, {Thomas},
  {Thompson}, {Kjeldseth-Moe}, {Brekke}, {Karud}, {Maltby}, {Aschenbach},
  {Br{\"a}uninger}, {K{\"u}hne}, {Hollandt}, {Siegmund}, {Huber}, {Gabriel},
  {Mason}, \& {Bromage}}]{Harrison-etal:95}
{Harrison}, R.~A., {Sawyer}, E.~C., {Carter}, M.~K., {et~al.} 1995, \solphys,
  162, 233

\bibitem[{{Hinteregger} {et~al.}(1981){Hinteregger}, {Fukui}, \&
  {Gilson}}]{Hinteregger-etal:81}
{Hinteregger}, H.~E., {Fukui}, K., \& {Gilson}, B.~R. 1981, \grl, 8, 1147

\bibitem[{{Jordan} \& {Brosius}(2007)}]{Jordan-Brosius:07}
{Jordan}, S.~D. \& {Brosius}, J.~W. 2007, in Astronomical Society of the
  Pacific Conference Series, Vol. 368, The Physics of Chromospheric Plasmas,
  ed. P.~{Heinzel}, I.~{Dorotovi{\v c}}, \& R.~J. {Rutten}, 183

\bibitem[{{Kamio} \& {Mariska}(2012)}]{Kamio-Mariska:12}
{Kamio}, S. \& {Mariska}, J.~T. 2012, \solphys, 279, 419

\bibitem[{{Mango} {et~al.}(1978){Mango}, {Bohlin}, {Glackin}, \&
  {Linsky}}]{Mango-etal:78}
{Mango}, S.~A., {Bohlin}, J.~D., {Glackin}, D.~L., \& {Linsky}, J.~L. 1978,
  \apj, 220, 683

\bibitem[{{Pauluhn} {et~al.}(2000){Pauluhn}, {Solanki}, {R{\"u}edi}, {Landi},
  \& {Sch{\"u}hle}}]{Pauluhn-etal:00}
{Pauluhn}, A., {Solanki}, S.~K., {R{\"u}edi}, I., {Landi}, E., \&
  {Sch{\"u}hle}, U. 2000, \aap, 362, 737

\bibitem[{{Pietarila} \& {Judge}(2004)}]{Pietarila-Judge:04}
{Pietarila}, A. \& {Judge}, P.~G. 2004, \apj, 606, 1239

\bibitem[{{Reeves}(1976)}]{Reeves:76}
{Reeves}, E.~M. 1976, \solphys, 46, 53

\bibitem[{{Richards} {et~al.}(2006){Richards}, {Woods}, \&
  {Peterson}}]{Richards-etal:06}
{Richards}, P.~G., {Woods}, T.~N., \& {Peterson}, W.~K. 2006, Advances in Space
  Research, 37, 315

\bibitem[{{Skumanich} {et~al.}(1975){Skumanich}, {Smythe}, \&
  {Frazier}}]{Skumanich-etal:75}
{Skumanich}, A., {Smythe}, C., \& {Frazier}, E.~N. 1975, \apj, 200, 747

\bibitem[{{Thompson} \& {Brekke}(2000)}]{Thompson-Brekke:00}
{Thompson}, W.~T. \& {Brekke}, P. 2000, \solphys, 195, 45

\bibitem[{{Tobiska} {et~al.}(2008){Tobiska}, {Bouwer}, \&
  {Bowman}}]{Tobiska-etal:08}
{Tobiska}, W.~K., {Bouwer}, S.~D., \& {Bowman}, B.~R. 2008, Journal of
  Atmospheric and Solar-Terrestrial Physics, 70, 803

\bibitem[{{Wang} {et~al.}(2011{\natexlab{a}}){Wang}, {Thomas}, {Brosius},
  {Young}, {Rabin}, {Davila}, \& {Del Zanna}}]{WangT-etal:11}
{Wang}, T., {Thomas}, R.~J., {Brosius}, J.~W., {et~al.} 2011{\natexlab{a}},
  \apjs, 197, 32

\bibitem[{{Wang} {et~al.}(2011{\natexlab{b}}){Wang}, {Robbrecht}, \&
  {Muglach}}]{WangYM-etal:11}
{Wang}, Y.-M., {Robbrecht}, E., \& {Muglach}, K. 2011{\natexlab{b}}, \apj, 733,
  20

\bibitem[{{Wilhelm} {et~al.}(1995){Wilhelm}, {Curdt}, {Marsch}, {Schuhle},
  {Lemaire}, {Gabriel}, {Vial}, {Grewing}, {Huber}, {Jordan}, {Poland},
  {Thomas}, {Kuhne}, {Timothy}, {Hassler}, \& {Siegmund}}]{Wilhelm-etal:95}
{Wilhelm}, K., {Curdt}, W., {Marsch}, E., {et~al.} 1995, \solphys, 162, 189

\bibitem[{{Wilhelm} {et~al.}(1998){Wilhelm}, {Lemaire}, {Dammasch}, {Hollandt},
  {Schuehle}, {Curdt}, {Kucera}, {Hassler}, \& {Huber}}]{Wilhelm-etal:98}
{Wilhelm}, K., {Lemaire}, P., {Dammasch}, I.~E., {et~al.} 1998, \aap, 334, 685

\bibitem[{{Woods} {et~al.}(2005){Woods}, {Eparvier}, {Bailey}, {Chamberlin},
  {Lean}, {Rottman}, {Solomon}, {Tobiska}, \& {Woodraska}}]{Woods-etal:05}
{Woods}, T.~N., {Eparvier}, F.~G., {Bailey}, S.~M., {et~al.} 2005, Journal of
  Geophysical Research (Space Physics), 110, 1312

\bibitem[{{Woods} {et~al.}(2012){Woods}, {Eparvier}, {Hock}, {Jones},
  {Woodraska}, {Judge}, {Didkovsky}, {Lean}, {Mariska}, {Warren}, {McMullin},
  {Chamberlin}, {Berthiaume}, {Bailey}, {Fuller-Rowell}, {Sojka}, {Tobiska}, \&
  {Viereck}}]{Woods-etal:12}
{Woods}, T.~N., {Eparvier}, F.~G., {Hock}, R., {et~al.} 2012, \solphys, 275,
  115

\end{thebibliography}
\end{document}